\documentclass[10pt,a4paper,twoside]{article}
\usepackage{RR}

\usepackage[latin1]{inputenc}
\usepackage{amsmath, amsthm, amssymb}
\usepackage{a4wide}
\usepackage{algorithmic}
\usepackage{algorithm}
\usepackage{subfigure}
\usepackage{xspace}
\usepackage{url}

\usepackage{figlatex}

\newtheorem{lemma}{Lemma}
\newtheorem{theorem}{Theorem}
\newtheorem{proposition}{Proposition}

\newtheorem{definition}{Definition}

\AtBeginDocument{}

\makeatletter
\newcommand{\eqnlabel}[1]{\xdef\@currentlabel{\theequation}\ltx@label{#1}}
\newcommand{\cnstslabel}[2][\thesubcounstraints]{
  \refstepcounter{subcounstraints}%
  \xdef\@currentlabel{#1}%
  \tagform@{#1}%
  \ltx@label{#2}\quad &}
\newcommand{\cnstlabelBIS}[2][\thesubcounstraints]{
  \xdef\@currentlabel{#1}%
  \ltx@label{#2}}
\newcounter{subcounstraints}[equation]

\newenvironment{LinearProgram}[2][Minimize]{%
  \begin{equation}%
    \let\label=\eqnlabel
    \begin{array}{l}%
      \textsc{#1 } #2, \\%
      \textsc{under the constraints}\\%
      \begin{cases}%
}{%
      \end{cases}
    \end{array}
  \end{equation}%
}

\newcommand{\EquationsNumbered}[1]{%
  \setcounter{subcounstraints}{0}
  \renewcommand{\thesubcounstraints}{\theequation\alph{subcounstraints}}
  \def\set@counstraintscounter{
    \refstepcounter{subcounstraints}%
    \xdef\@currentlabel{\thesubcounstraints}%
    \tagform@{\thesubcounstraints}%
  }
  \def\mark{\set@counstraintscounter\quad&}
  \def\n{\\\mark}
  \begin{aligned}%
    #1
  \end{aligned}%
}
\makeatother

\newcommand{\greedyalgo}{\textsc{Best-Balance Algorithm}\xspace}
\newcommand{\greedy}{BBA\xspace}
\newcommand{\moorealgo}{\textsc{Moore Based Binary-Search Algorithm}\xspace}
\newcommand{\moore}{MBBSA\xspace}
\newcommand{\backwardalgo}{\textsc{Reversed Binary-Search
    Algorithm}\xspace}
\newcommand{\backward}{R-BSA\xspace}

\def\accolade#1{$\left\{\begin{array}{c}\vspace{#1}\end{array}\right.$}%

\RRetitle{Scheduling and data redistribution strategies on star platforms}

\RRtitle{Strat\'egies d'ordonnancement et de redistribution de donn\'ees sur
  plate-formes en \'etoile}

\newboolean{oncover}
\setboolean{oncover}{true}

\newcounter{alreadyused}
\RRdate{\ifthenelse{\equal{\value{alreadyused}}{0}}{%
October 2006\\[10mm]%
\begin{tabular}{c}%
\text{LIP UMR 5668 CNRS-ENS Lyon-INRIA UCBL}\\[1mm]
\text{Research Report N\textsuperscript{o} 2006-23}
\end{tabular}}{October 2006}\stepcounter{alreadyused}}

\RRauthor{Loris Marchal\and Veronika Rehn\and Yves Robert\and Fr\'ed\'eric Vivien}
\titlehead{Scheduling and data redistribution strategies on star platforms}
\authorhead{L. Marchal, V. Rehn, Y. Robert and  F. Vivien}

\RRabstract{
  In this work we are interested in the problem of scheduling and redistributing
  data on master-slave platforms. We consider the case were the workers possess
  initial loads, some of which having to be redistributed in order to balance their completion times.
  
  We examine two different scenarios. The first model assumes that the data consists of independent
  and identical tasks. We prove the NP-completeness in the strong sense for the
  general case,  and we present two optimal algorithms for special platform
  types. Furthermore we propose three heuristics for the general
  case. Simulations consolidate the theoretical results.
  
  The second data model is based on Divisible Load Theory. This problem can be
  solved in polynomial time by a combination of linear programming and
  simple analytical manipulations.
}

\RRkeyword{Master-slave platform, scheduling, data redistribution, 
  one-port model, independent tasks, divisible load theory.}

\RRresume{
Dans ce travail on s'interesse au probl\`eme d'ordonnancement et de
redistribution de donn\'ees sur plates-formes ma\^itre-esclaves. On consid\`ere le
cas o\`u les esclaves poss\`edent des donn\'ees initiales, dont quelques-unes
doivent \^etre redistribu\'ees pour \'equilibrer leur dates de fin.

On examine deux sc\'enarios diff\'erents. Le premier mod\`ele suppose que les
  donn\'ees sont des t\^aches ind\'ependantes identiques. On prouve
 la NP-compl\'etude dans le sens fort pour le cas g\'en\'eral, et on pr\'esente
  deux algorithmes pour des plates-formes sp\'eciales. De plus on propose trois
  heuristiques pour le cas g\'en\'eral. Des r\'esultats exp\'erimentaux obtenus
  par simulation viennent \`a l'appui des
  r\'esultats th\'eoriques.
  }

\RRmotcle{Plate-forme ma\^itre-esclave, ordonnancement, \'equilibrage de
  charge, mod\`ele un-port, t\^aches ind\'ependantes, t\^aches divisibles.}
\RRtheme{\THNum}
\URRhoneAlpes
\RRprojet{GRAAL}

\begin{document}
\makeRR

\tableofcontents
\newpage

\section{Introduction}


In this work we consider the problem of scheduling and redistributing data on
master-slave architectures in star topologies. Because of
variations in the resource performance (CPU speed or communication
bandwidth), or because of unbalanced amounts of current load on the workers, data must be
redistributed between the participating processors,  so that the updated load is
better balanced in terms that the overall processing finishes earlier.

We adopt the following abstract view of our problem. There are $m+1$
participating processors $P_0, P_1, \dots, P_m$, where $P_0$ is the
master. Each processor $P_k$, $1\leq k \leq m$ initially holds $L_k$ data
items. During our scheduling process we try to determine which processor $P_i$
should send some data to another worker $P_j$ to equilibrate their
finishing times. The goal is to minimize the global makespan, that is the time
until each processor has finished to process its data.
Furthermore we suppose that each communication link is fully bidirectional,
with the same bandwidth for
receptions and sendings. This assumption is quite realistic in practice, and does not
change the complexity of the scheduling problem, which we prove NP-complete in the strong sense.

We examine two different scenarios for the data items that are situated at the
workers. The first model supposes that these data items consist in independent and
uniform tasks, while the other model uses the \textsc{Divisible Load Theory} paradigm
(DLT) \cite{Bharadwaj-Cluster2003}.

The core of DLT is the following: DLT assumes that communication and computation
loads can be fragmented into parts of arbitrary size and then distributed arbitrarily among different processors
to be processed there. This corresponds to perfect parallel jobs: They can be split into arbitrary subtasks which can be processed in
parallel in any order on any number of processors. 

Beaumont, Marchal, and Robert~\cite{BeauMarRob} treat the problem of divisible loads with return messages on
heterogeneous master-worker platforms (star networks). In their framework, all the initial load is
situated at the master and then has to be distributed to the
workers. The workers compute their amount of load and return their results to the
master. The difficulty of the problem is to decide about the sending order
from the master and, at the same time, about the receiving order. In this paper
problems are formulated in terms of linear programs. Using this approach the authors
were able to characterize optimal LIFO\footnote{Last In First Out} and
FIFO\footnote{First In First Out} strategies, whereas the general
case is still open.
Our problem is different, as in our case the initial load is
already situated at the workers. To the best of our knowledge, we are the first
to tackle this kind of problem.

Having discussed the reasons and background of DLT,
we dwell on the interest of the data model with uniform and independent
tasks. Contrary to the DLT model, where the size of load can be diversified, the
size of the tasks has to be fixed at the beginning. This leads to the
first point of interest: When tasks have different sizes, the problem is
NP complete because of an obvious reduction to 2-partition \cite{GareyJohnson}.
The other point is a positive one: there
exists lots of practical applications who use fixed identical and independent
tasks. A famous example is BOINC \cite{BOINC}, the Berkeley Open Infrastructure for
Network Computing, an open-source software
platform for volunteer computing. It works as a centralized scheduler that
distributes tasks for participating applications. These projects consists in
the treatment of computation extensive and expensive scientific problems of
multiple domains, such as biology, chemistry or mathematics. SETI@home \cite{SETI} for
example uses the accumulated computation power for the search of
extraterrestrial intelligence. In the astrophysical domain, Einstein@home
\cite{EINSTEIN} searches for spinning neutron stars using data from the LIGO
and GEO gravitational wave detectors. To get an idea of the task dimensions,
in this project a
task is about 12~MB and requires between 5 and 24 hours of dedicated computation.

As already mentioned, we suppose that all data are initially situated on the
workers, which leads us to a kind of redistribution problem.
Existing redistribution algorithms have a different objective. Neither do they care how the
degree of imbalance is determined, nor do they include the computation phase
in their optimizations. They expect that a load-balancing algorithm has
already taken place. With help of these results, a redistribution algorithm
determines the required communications and organizes them in minimal
time. Renard, Robert, and Vivien present some optimal redistribution algorithms
for heterogeneous processor rings in~\cite{rr2004-redistrib}.  We
could use this approach and redistribute the data first and then enter in a
computation phase. But our
problem is more complicated as we suppose that communication and computation can
overlap, i.e., every worker can start computing its initial data while the
redistribution process takes place.

To summarize our problem: as the participating workers are not equally charged
and/or because of different resource performance, they might not finish their
computation process at the same time. So we are looking for mechanisms on how to
redistribute the loads in order to finish the global computation process in minimal
time under the hypothesis that charged workers can compute at the same time as
they communicate.

The rest of this report is organized as follows: Section~\ref{sec:state}
presents some related work. The data model
of independent and identical tasks is treated in Section~\ref{sec:tasks}:
In Section~\ref{sec:general} we discuss the case of
general platforms. We are able to prove the NP-completeness for the general case
of our problem,  and the polynomiality for a restricted problem. The following
sections consider some particular platforms:
 an optimal algorithm for homogeneous star
networks is presented in Section~\ref{sec:greedy}, Section
\ref{sec:moore} treats platforms with homogenous communication links and
heterogeneous workers.  The presentation of some heuristics for heterogeneous
platforms is the subject in Section~\ref{sec:heuristic}. Simulative test results are
shown in Section~\ref{sec:tests}. Section~\ref{sec:divisible} is
devoted to the DLT model. We propose a linear program to solve the scheduling
problem and propose formulas for the redistribution process.

\section{Related work}
\label{sec:state}
Our work is principally related with three key topics.
Since the early nineties \textsc{Divisible Load Theory (DLT)} has been assessed to be an interesting method of
distributing load
in parallel computer systems. The outcome of DLT is a huge variety of scheduling strategies on how to
distribute the independent parts to achieve maximal results. As the DLT model can be used on a vast variety of interconnection topologies like trees, buses, hypercubes and so on, in the
literature theoretical and applicative elements are widely
discussed. In his article Robertazzi gives
{\it Ten Reasons to Use Divisible Load Theory}
\cite{Robertazzi-Computer2003}, like scalability or extending realism. Probing strategies \cite{Ghose02} were
shown to be able to handle unknown platform parameters.
In \cite{DrozWiel} evaluations of efficiency of DLT are conducted. The authors analyzed the relation between the values of particular
parameters and the efficiency of parallel computations. They demonstrated that  several parameters in parallel systems are mutually
related, i.e., the change of one of these
parameters should be accompanied by the changes of the other parameters to keep efficiency. The platform used in this article is a star
network and the results are for applications with no return messages.
Optimal scheduling algorithms including
return messages are presented in \cite{AltilarPak02}. The authors are treating the problem of processing digital video
sequences for digital TV and interactive multimedia. As a result, they propose two optimal algorithms for real time frame-by-frame
processing.
Scheduling problems with multiple sources are examined \cite{MogesYu}. The authors
propose closed form solutions for tree networks with two load originating processors.

\textsc{Redistribution algorithms} have also been well studied in the
literature. Unfortunately already simple redistribution problems are NP
complete \cite{kremer93npcompleteness}. For this reason, optimal algorithms can be designed only for
particular cases, as it is done in \cite{rr2004-redistrib}. In their research,
the authors restrict the platform architecture to ring topologies, both
uni-directional and bidirectional. In the homogeneous case, they were able to
prove optimality, but the heterogenous case is still an open problem.
  In spite of this, other
efficient algorithms have been proposed. For topologies like trees or
hypercubes some results are
presented in \cite{Wu97}.

The \textsc{load balancing problem} is not directly dealt with in this paper. Anyway we want to
quote some key references to this subject, as the results of these algorithms
are the starting point for the redistribution process. Generally load
balancing techniques can be classified into two categories. Dynamic load
balancing strategies and static load balancing.  Dynamic techniques might use the past for
the prediction of the future as it is the case in \cite{CiernakZakiLi97b} or they suppose that
the load varies permanently \cite{HamdiLee95}. That is why for our problem static
algorithms are more interesting: we are only treating star-platforms and as
the amount of load to be treated is known a priory we do not need
prediction. For homogeneous platforms, the
papers in \cite{ieee-sched} survey existing results. Heterogeneous solutions are
presented in \cite{NibhanupudiSzymanski} or \cite{Informatica}. This last paper is about
a dynamic load balancing method for data parallel applications, called the
\textsc{working-manager method}: the manager is supposed to use its idle time
to process data itself. So the heuristic is simple: when the manager does not
perform any control task it has to work, otherwise it schedules.

\section{Load balancing of independent tasks using the one-port bidirectional model}
\label{sec:tasks}

\subsection{Framework}
\label{sec:framework_tasks}

\begin{figure}[htbp]
    \centering
    \includegraphics[width=0.45\textwidth]{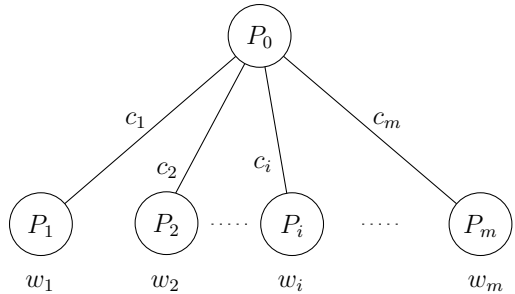}%
    \caption{Example of a star network.}
    \label{fig:star}
  \end{figure}

In this part we will work with a {\it star network} $S = {P_{0}, P_{1}, \ldots, P_{m}}$ shown in Figure~\ref{fig:star}. The processor $P_{0}$ is the
master and the $m$ remaining processors $P_{i}$, $1 \leq i \leq m$, are
workers. The initial data are distributed on the workers, so every
worker $P_i$ possesses a number $L_i$ of initial tasks. All tasks are
independent and identical. As we assume a linear cost model, each worker
$P_{i}$ has a (relative) computing power $w_{i}$ for the computation of one task: it takes $X.w_{i}$ time units to execute $X$ tasks on the worker
$P_{i}$. The master $P_{0}$ can communicate with each worker $P_{i}$ via a
communication link.  A worker $P_i$ can send
some tasks via the master to another worker $P_j$ to decrement its execution
time.  It takes $X.c_{i}$ time units to
send $X$ units of load from $P_{i}$ to $P_{0}$ and $X.c_{j}$ time units to
send these $X$ units from $P_{0}$ to a worker $P_{j}$. Without loss of
generality we assume that the master is not computing, and only communicating.

The platforms dealt with in sections~\ref{sec:greedy} and~\ref{sec:moore} are a special case of a star network: all communication links have the same
characteristics, i.e., $c_{i} = c$ for each processor $P_{i}$, $1 \leq i \leq k$. Such a platform is called a
{\it bus network} as it has homogeneous communication links.

We use the bidirectional one-port model for communication. This means, that
the master can only send data to, and receive data from,
a single worker at a given time-step. But it can simultaneously receive a data
and send one. A given worker cannot start an execution
before it has terminated the reception of the message from the master;
similarly, it cannot
start sending the results back to the master before finishing the computation.

The objective function is to minimize the makespan, that is the time at which
all loads have been processed. So we look for a schedule $\sigma$ that
accomplishes our objective.

\subsection{General platforms}
\label{sec:general}
Using the notations and the platform topology introduced in Section
\ref{sec:framework_tasks}, we now formally present the \textsc{Scheduling
  Problem for Master-Slave Tasks
 on a Star of Heterogeneous Processors} (SPMSTSHP).
\newpage
\begin{definition}[SPMSTSHP]
~

Let $N$ be a star-network with one special processor $P_0$ called ``master" and
$m$ workers. Let $n$ be the number
of identical tasks distributed to the workers.
For each
worker $P_i$, let $w_i$ be the computation time for one task. Each
communication link, $link_i$, has an associated communication time $c_i$ for the
transmission of one task. Finally let $T$ be a deadline.

The question associated to the decision problem of SPMSTSHP is: ``Is it
possible to redistribute the tasks and to process them in time $T$?''.
\end{definition}

One of the main difficulties seems
to be the fact that we cannot partition the workers into disjoint sets of
senders and receivers. There exists situations where, to minimize the global makespan, it
is useful, that sending workers also receive tasks. (You will see later in
this report that we can suppose this distinction when communications are
homogeneous.)

We consider the
following example. We have four workers (see Figure~\ref{tab:param} for their
parameters) and a makespan fixed to $M = 12$.
An optimal solution is shown in Figure~\ref{fig:contre_exemple}: Workers $P_3$ and $P_4$ do not own any task, and
they are computing very slowly. So each of them can compute exactly one task. Worker $P_1$, who is a fast processor and
communicator, sends them their tasks and receives later another task from
worker $P_2$ that it can
compute just in time. Note that worker $P_1$ is both sending and
receiving tasks. Trying to solve the problem under the constraint that no worker
also sends and receives, it is not feasible to achieve a makespan of
12. Worker $P_2$ has to send one task either to worker $P_3$ or to worker
$P_4$. Sending and receiving this
task takes 9 time units. Consequently the processing of this task can not finish earlier than
time $t=18$.

\begin{figure}
  \begin{minipage}{0.4\textwidth}
    \centering
    \begin{tabular}{|l|c c c|}
      \hline
      Worker & c & w & load \\
      \hline
      $P_1$ & 1 & 1 & 13\\
      $P_2$ & 8 & 1 & 13\\
      $P_3$ & 1 & 9 & 0\\
      $P_4$ & 1 & 10 & 0\\
      \hline
    \end{tabular}
    \caption{Platform parameters.}
    \label{tab:param}
  \end{minipage}\hfill
  \begin{minipage}{0.5\textwidth}
    \centering
    \includegraphics[width=0.8\textwidth]{contreExempleHet.fig}%
    \caption{Example of an optimal schedule on a heterogeneous platform, where a
      sending worker also receives a task.}
    \label{fig:contre_exemple}
  \end{minipage}
\end{figure}

Another difficulty of the problem is the overlap of computation and the
redistribution process. Subsequently we examine our problem neglecting the computations. We are going to prove an optimal
polynomial algorithm for this problem.

\subsubsection{Polynomiality when computations are neglected}
Examining our original problem under the supposition that computations are negligible, we get a
classical data redistribution problem. Hence we eliminate the original
difficulty of the
overlap of computation  with the data redistribution process. We suppose that we already know the imbalance of the
system.
So we adopt the following abstract view of our new problem: the $m$
participating workers $P_1, P_2, \dots P_m$ hold their initial uniform tasks
$L_i$, $1\leq i \leq m$. For a worker $P_i$ the chosen
algorithm for the computation of the imbalance has decided that the new load should be $L_i - \delta_i$. If
$\delta_i > 0$, this means that $P_i$ is overloaded and it has to send
$\delta_i$ tasks to some other processors. If $\delta_i < 0$, $P_i$ is underloaded
and it has to receive $-\delta_i$ tasks from other workers. We have
heterogeneous communication links and all sent tasks pass by the master. So
the goal is to determine the order of senders and receivers to redistribute
the tasks in minimal time.

As all communications pass by the master, workers can not start receiving until
tasks have arrived on the master. So to minimize the redistribution time, it
is important to charge the master as fast as possible. Ordering the senders by
non-decreasing $c_i$-values makes the tasks at the earliest possible time available.

\begin{figure}[htbp]
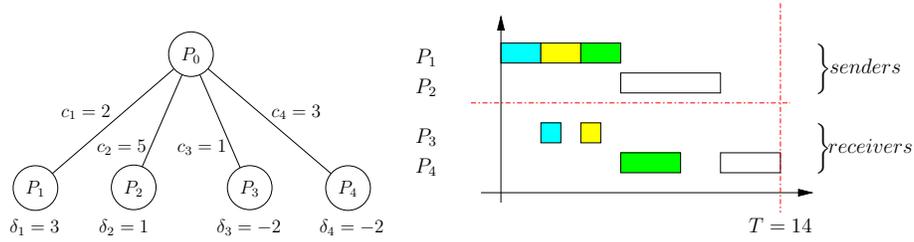
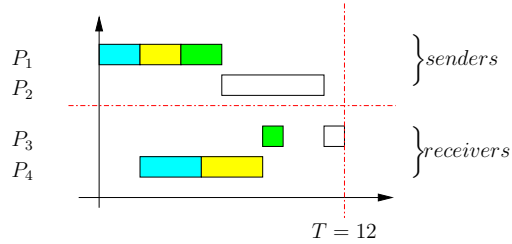

    \centering
    \subfigure[Example of load imbalance on a heterogeneous platform with 4 workers.]{
      \includegraphics[width=0.32\textwidth]{platformHetTheorem.fig}%
      \label{fig:platHet}%
    } $\quad$
    \subfigure[The receivers are ordered by non-decreasing order of their
    $c_i$-values.]{
      \includegraphics[width=0.4\textwidth]{fastFirst.fig}%
      \label{fig:fastFirst}%
    } $\quad$
    \subfigure[The receivers are ordered by non-increasing order of their
    $c_i$-values.]{%
      \includegraphics[width=0.4\textwidth]{slowFirst.fig}%
      \label{fig:slowFirst}%
    }
    \caption{Comparison of the ordering of the receivers.}
  \end{figure}

 Suppose we would order the receivers in the same manner
as the senders, i.e., by non-decreasing $c_i$-values. In this case we could start each
reception as soon as possible, but always with the restriction that each task
has to arrive first at the master (see Figure~\ref{fig:fastFirst}). So it can
happen that there are many idle times between the receptions if the tasks do
not arrive in time on the master. That is why we choose to order the receiver
in reversed order, i.e.,  by non-increasing $c_i$-values (cf. Figure
\ref{fig:slowFirst}), to let the tasks more time to arrive. In the following lemma we even prove optimality of this ordering.

\begin{theorem}
\label{th:heterogeneous_order}
Knowing the imbalance $\delta_i$ of each processor, an optimal solution
for heterogeneous star-platforms
is to order the senders by non-decreasing $c_i$-values and the receivers by
non-increasing order of $c_i$-values.
\end{theorem}

\begin{proof}
To prove that the scheme described by Theorem~\ref{th:heterogeneous_order} returns an optimal
schedule, we take a schedule $S'$ computed by this scheme. Then we take
any other schedule $S$. We are going to transform $S$ in two steps into our schedule
$S'$ and prove that the makespans of the both schedules hold the following
inequality: $M(S') \leq M(S)$.

In the first step we take a look at the senders. The sending from the master can not start before tasks are
available on the master. We do not know the ordering
of the senders in $S$ but we know the ordering in $S'$: all senders are
ordered in non-decreasing order of their $c_i$-values. Let $i_0$ be the first
task sent in $S$ where the sender of task $i_0$ has a bigger $c_i$-value than
the sender of the $(i_0+1)$-th task. We then exchange the senders of task $i_0$ and task $(i_0 + 1)$ and call this new schedule $S_{new}$. Obviously
the reception time for the second task is still the same. But as you
can see in Figure~\ref{fig:sender}, the time when the first task is available
on the master has changed: after the exchange, the first task is available
earlier and ditto ready for reception. Hence this exchange improves the
availability on the master (and reduces possible idle times for the
receivers). We use this mechanism to transform the sending order of $S$ in
the sending order of $S'$ and at each time the availability on the master is
improved. Hence at the end of the transformation the makespan of $S_{new}$
is smaller than or equal to that of $S$ and the sending order of $S_{new}$ and $S'$ is the
same.

 \begin{figure}[htb]
   \centering
  \includegraphics[width=0.7\textwidth]{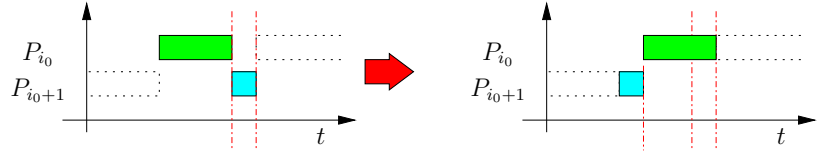}%
  \caption{Exchange of the sending order makes tasks available earlier on the master.}
  \label{fig:sender}
\end{figure}

In the second step of the transformation we take care of the receivers (cf. Figures
\ref{fig:receiver} and~\ref{fig:receiver_idle}). Having
already changed the sending order of $S$ by the first transformation of $S$
into $S_{new}$, we start here directly by the transformation of $S_{new}$. Using
the same mechanism as for the senders, we call $j_0$ the first task such that the
receiver of task $j_0$ has a smaller $c_i$-value than the receiver of task
$j_0+1$. We exchange the receivers of the tasks $j_0$ and $j_0+1$ and call
the new schedule $S_{new^{(1)}}$.
$j_0$ is sent at the same time than previously, and the processor receiving
it, receives it earlier than it received $j_{0+1}$ in $S_{new}$.
$j_{0+1}$ is sent as soon as it is available on the master and as soon as the
communication of task $j_0$ is completed. The first of these two conditions
had also to be satisfied by $S_{new}$. If the second condition is delaying the
beginning of the sending of the task $j_0+1$ from the master, then this
communication ends at time $t_{in}+c_{\pi' (j_0)} + c_{\pi' (j_0+1)} =
t_{in}+c_{\pi (j_0+1)} + c_{\pi (j_0)}$ and this communication ends at the same
time than under the schedule $S_{new}$ ( here $\pi (j_0)$ ($\pi' (j_0)$) denotes the receiver of
task $j_0$ in schedule $S_{new}$ ($S_{new^{(1)}}$, respectively)). Hence the finish time of the
communication of task $j_0+1$  in schedule $S_{new^{(1)}}$ is less than or
equal to the finish time in the previous schedule. In all cases, $M(S_{new^{(1)}})\leq
M(S_{new})$. Note that this transformation does not change anything for the
tasks received after $j_{0+1}$ except that we always perform the scheduled
communications as soon as possible.
Repeating the
transformation for the rest of the schedule $S_{new}$ we reduce all idle times
in the receptions as far as possible. We get for the makespan of each schedule $S_{new^{(k)}}$:
$M(S_{new^{(k)}}) \leq M(S_{new}) \leq M(S)$. As after these (finite number of) transformations the
 order of the receivers will be in non-decreasing order of the $c_i$-values,
 the receiver order of $S_{new^{(\infty)}}$ is the same as the receiver order of $S'$
 and hence we have $S_{new^{(\infty)}} = S'$. Finally we conclude that the makespan of
 $S'$ is smaller than or equal to any other schedule $S$ and hence $S'$ is optimal.

 \begin{figure}[htb]
  \centering
  \includegraphics[width=0.7\textwidth]{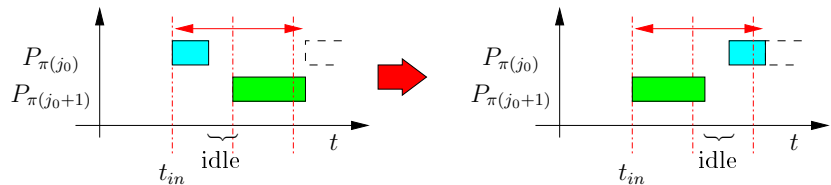}%
  \caption{Exchange of the receiving order suits better with the available
    tasks on the master.}
  \label{fig:receiver}
\end{figure}

 \begin{figure}[htb]
  \centering
  \includegraphics[width=0.7\textwidth]{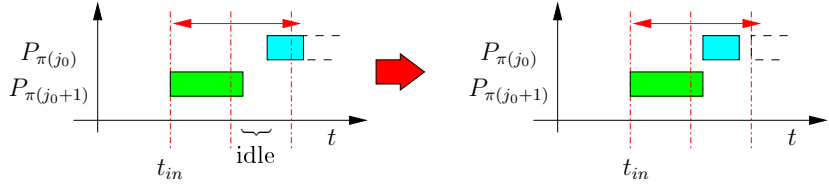}%
  \caption{Deletion of idle time due to the exchange of the receiving order.}
  \label{fig:receiver_idle}
\end{figure}

\end{proof}

\subsubsection{NP-completeness of the original problem}
Now we are going to prove the NP-completeness in the strong sense of the
general problem. For this we were strongly inspired by the proof of Dutot
\cite{ejorDutot05, PhD-Dutot_04} for the \textsc{Scheduling Problem for Master-Slave Tasks
 on a Tree of Heterogeneous Processors} (SPMSTTHP). This proof uses a two level tree as platform topology and we
are able to associate the structure on our star-platform. We are going to recall the
3-partition problem which is NP-complete in the strong sense
\cite{GareyJohnson}.

\begin{definition}[3-Partition]~

Let S and n be two integers, and let $(y_i)_{i\in 1..3n}$ be a sequence of $3n$
integers such that for each $i$, $\frac{S}{4}<y_i<\frac{S}{2}$.

The question of the 3-partition problem is ``Can we partition the set of the
$y_i$ in $n$ triples such that the sum of each triple is exactly $S$?".
\end{definition}

\begin{theorem}
SPMSTSHP is NP-complete in the strong sense.
\end{theorem}

\begin{proof}
We take an instance of 3-partition. We define some real numbers $x_i$, $1\leq
i \leq 3n$,  by $x_i =
\frac{1}{4}S + \frac{y_i}{8}$. If a triple of $y_i$ has the sum $S$, the corresponding triple of $x_i$
corresponds to the sum $\frac{7S}{8}$ and vice versa. A partition of $y_i$ in
triples is thus equivalent to a partition of the $x_i$ in triples of the sum
$\frac{7S}{8}$.
This modification allows us to guarantee that the $x_i$ are contained in a
smaller interval than the interval of the $y_i$. Effectively the $x_i$ are
strictly included between $\frac{9S}{32}$ and $\frac{5S}{16}$.

\paragraph{Reduction.}
For our reduction we use the star-network shown in Figure
\ref{fig:star_reduction}. We consider the following instance of SPMTSHP:
Worker $P$ owns $4n$ tasks, the other $4n$ workers do not hold any
task. We work with the deadline $T = E +nS + \frac{S}{4}$, where $E$ is an
enormous time fixed to $E=(n+1)S$. The communication link
between $P$ and the master has a $c$-value of $\frac{S}{4}$. So it can send a task
all $\frac{S}{4}$ time units. Its computation time is $T+1$, so worker $P$ has to
distribute all its tasks as it can not finish processing a single task by the deadline. Each of
the other workers is able to process one single task, as its computation time
is at least $E$ and we have $2E >T$, what makes it impossible to process a
second task by the deadline.

 \begin{figure}[htb]
   \centering
  \includegraphics[width=0.9\textwidth]{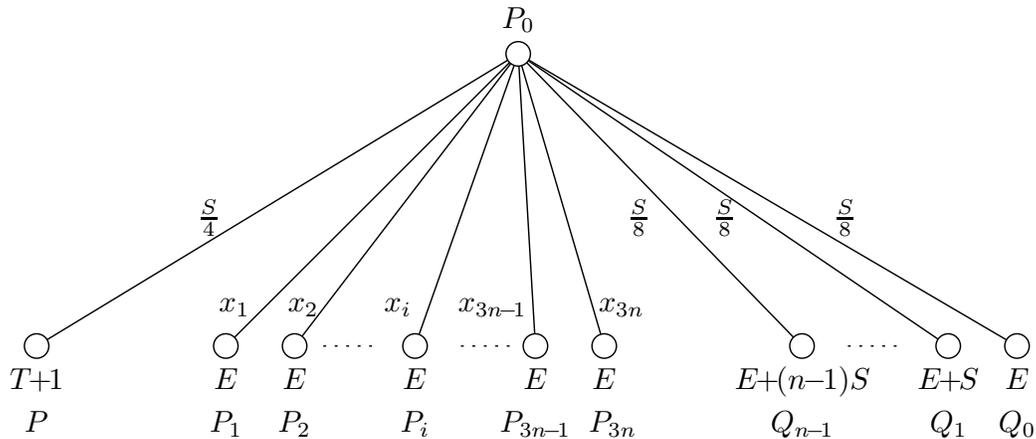}%
  \caption{Star platform used in the reduction.}
  \label{fig:star_reduction}
\end{figure}

This structure of the star-network is particularly constructed to reproduce
the 3-partition problem in the scope of a scheduling problem. We are going to
use the bidirectional 1-port constraint to create our triplets.

\paragraph{Creation of a schedule out of a solution to 3-partition.}
First we show how to construct a valid schedule of $4n$ tasks in time
$\frac{S}{4} + nS + E$ out of a 3-partition solution. To facilitate the lecture, the processors $P_i$ are
ordered by their $x_i$-values in the order that corresponds to the solution of 3-partition.
So, without loss of
generality, we assume that for each $j \in [0,n-1]$, $x_{3j+1} + x_{3j+2} +
x_{3j+3} = \frac{7S}{8}$.
The schedule is of the following form:
\begin{enumerate}
\item
Worker $P$ sends its tasks as soon as possible to the master, i.e., every $\frac{S}{4}$ time
units. So it is guaranteed that the $4n$ tasks are sent in $nS$ time units.

\item
The master sends the tasks as soon as possible in incoming order to the
workers. The receiver order is the following (for all $j \in [0,n-1]$):
\begin{itemize}
\item
Task $4j+1$, over link of cost $x_{3j+1}$, to processor $P_{3j+1}$.
\item
Task $4j+2$, over link of cost $x_{3j+2}$, to processor $P_{3j+2}$.
\item
Task $4j+3$, over link of cost $x_{3j+3}$, to processor $P_{3j+3}$.
\item
Task $4j+4$, over link of cost $\frac{S}{8}$, to processor $Q_{n-1-j}$.
\end{itemize}
\end{enumerate}

The distribution of the four tasks, $4j+1$, $4j+2$, $4j+3$, $4j+4$, takes
exactly $S$ time units and the master needs
also $S$ time units to receive four tasks from processor $P$. Furthermore,
each $x_i$ is larger than $\frac{S}{4}$. Therefore, after the first task is
sent, the master always finishes to receive a new task before its outgoing
port is available to send it. The first task arrives
at time $\frac{S}{4}$ at the master, which is responsible for the short idle
time at the beginning. The last task arrives at its worker at time
$\frac{S}{4}+nS$ and hence it rests exactly $E$ time units for the
processing of this task. For the workers $P_i$, $1 \leq i \leq 3n$, we know that they can
finish to process their tasks in time as they all have a computation power of
$E$. The computation power of the workers $Q_i$, $0 \leq i \leq n-1$, is $E +
i\times S$ and as they receive their task at time $\frac{S}{4} + (n-i-1)\times
S + \frac{7S}{8}$, they have exactly the time
to finish their task.

\paragraph{Getting a solution for 3-partition out of a schedule.}
 Now we prove that each schedule of $4n$ tasks in time $T$ creates a solution
 to the 3-partition problem.

As already mentioned, each worker besides  worker $P$ can process at most one
task. Hence due to the number of tasks in the system, every worker has to
process exactly one task. Furthermore the minimal time needed to distribute
all tasks from the master and the minimal processing time on the workers
induces that there is no idle time in the emissions of the master, otherwise
the schedule would take longer than time $T$.

We also know that worker $P$ is the only sending worker:

\begin{lemma}
No worker besides worker $P$ sends any task.
\end{lemma}

\begin{proof}
Due to the platform configuration and the total number of tasks, worker $P$
has to send all its tasks. This takes at least $nS$ time units. The total
emission time for the master is also $nS$ time units: as each worker must
process a task, each of them must receive one. So the emission time for the
master is larger than or equal to $\sum_{i=1}^{n}x_i + n\times \frac{S}{8} =
nS$. As the master cannot start sending the first task before time
$\frac{S}{4}$ and as the minimum computation power is $E$, then if the master
sends exactly one task to each slave, the makespan is greater than or equal to
$T$ and if one
worker besides $P$ sends a task, the master will at least send one additional
task and the makespan will be strictly greater than $T$.
\end{proof}

Now we are going to examine the worker $Q_{n-1}$ and the task he is
associated to.

\begin{lemma}
The task associated to worker $Q_{n-1}$ is one of the first four tasks sent by
worker $P$.
\end{lemma}

\begin{proof}
The computation time of worker $Q_{n-1}$ is $E + (n-1)S$, hence its task has to
arrive no later than time $S+ \frac{S}{4}$. The fifth task arrives at the soonest at
time $\frac{5S}{4} + \frac{S}{8}$ as  worker $P$ has to send five tasks as the
shortest communication time is $\frac{S}{8}$. The following tasks arrive
later than the $5$-th task, so the task for worker $Q_{n-1}$ has to be one of
the first four tasks.
\end{proof}

\begin{lemma}
The first three tasks are sent to some worker $P_i$, $1 \leq i \leq 3n$.
\end{lemma}

\begin{proof}
As already mentioned, the master has to send without any idle time besides the
initial one. Hence we have to pay attention that the master always possesses a
task to send when he finishes to send a task. While the master is sending to a
worker $P_i$, worker $P$ has the time to send the next task to the
master. But, if at least one of the first three tasks is sent to a worker
$Q_i$, the sending time of the first three tasks is strictly inferior to
$\frac{S}{8}+\frac{5}{16}S+\frac{5}{16}S=\frac{3}{4}S$. Hence there is
obligatory an idle time in the emission of the master. This pause makes the
schedule of $4n$ tasks in time $T$ infeasible.
\end{proof}

A direct conclusion of the two precedent lemmas is that the $4$-th task is
sent to worker $Q_{n-1}$.

\begin{lemma}
The first three tasks sent by worker $P$ have a total communication time of
$\frac{7}{8}S$ time units.
\end{lemma}

\begin{proof}
Worker $Q_{n-1}$ has a computation time of $E+(n-1)S$, it has to receive its
task no later than time $\frac{5}{4}S$. This implies that the first three
tasks are sent in a time no longer than $\frac{7}{8}S$.

On the other side, the $5$-th task arrives at the master no sooner than time
$\frac{5}{4}S$. As the master has to send without idle time, the emission to
worker $Q_{n-1}$ has to persist until this date. Necessarily the first three
emissions of the master take at minimum a time $\frac{7}{8}S$.
\end{proof}

\begin{lemma}
Scheduling $4n$ tasks in a time $T = \frac{S}{4}+ nS + E$ units of time allows
to reconstruct an instance of the associated 3-partition problem.
\end{lemma}

\begin{proof}
In what precedes, we proved that the first three tasks sent by the master
create a triple whose sum is exactly $\frac{7}{8}$. Using this property
recursively on $j$ for the triple $4j+1,\; 4j+2$ and $4j+3$, we show that we
must send the tasks
$4j+4$ to the worker $Q_{n-1-j}$. With this method we construct a partition of
  the set of $x_i$ in triples of sum $\frac{7}{8}$. These triples are a
  solution to the associated 3-partition problem.
\end{proof}

Having proven that we can create a schedule out of a solution of 3-partition
and also that we can get a solution for 3-partition out of a schedule, the
proof is now complete.

\end{proof}

\subsection{An algorithm for scheduling on homogeneous star platforms: the
  best-balance algorithm}
\label{sec:greedy}

In this section we present the \greedyalgo (\greedy), an
algorithm to schedule on homogeneous star platforms. As already mentioned,
we use a bus network with communication speed $c$, but additionally we
suppose that the
computation powers are homogeneous as well. So we have $w_i = w$ for all $i$,
$1\leq i \leq m$.

The idea of \greedy is simple: in each iteration, we look if we could finish earlier if we
redistribute a task. If so, we schedule the task, if not, we stop redistributing. The algorithm has polynomial run-time. It is a
natural intuition that \greedy is optimal on homogeneous platforms, but the
formal proof is rather complicated, as can be seen in Section~\ref{sec:star_hom}.

\subsubsection{Notations used in \greedy}
\greedy schedules one task per iteration $i$. Let
  $L_{k}^{(i)}$ denote the number of tasks of worker $k$ after iteration
  $i$, i.e., after $i$ tasks were redistributed. The date at which the master has finished receiving the $i$-th task is
  denoted by $master\_in^{(i)}$. In the same way we call $master\_out^{(i)}$ the date at which the master has finished sending the $i$-th task.
Let $end_{k}^{(i)}$ be the date at which worker $k$ would finish to process
  the load it would hold if exactly
  $i$ tasks are redistributed. The worker $k$ in iteration $i$ with the biggest finish time
$end_{k}^{(i)}$, who is chosen to send one task in the next iteration, is
  called $sender$. We call $receiver$ the worker $k$ with smallest finish time $end_{k}^{(i)}$ in
iteration $i$ who is chosen to receive one task in the next iteration.

In iteration $i = 0$ we are in the initial configuration:
All workers own their initial tasks $L_{k}^{(0)} = L_k$ and the makespan of each worker $k$ is
the time it needs to compute all its tasks: $end_{k}^{(0)} = L_{k}^{(0)} \times
w$. $master\_in^{(0)} = master\_out^{(0)} = 0$.

\subsubsection{The Best Balance Algorithm - BBA}
\label{sec:star_hom}

We first sketch \greedy:\\
In each iteration $i$ do:
\begin{itemize}
\item
  Compute the time $end_{k}^{(i-1)}$ it would take worker $k$ to process $L_{k}^{(i-1)}$ tasks.
\item
  A worker with the biggest finish time $end_{k}^{(i-1)}$ is arbitrarily
  chosen as sender, he is called
  $sender$.
\item
  Compute the temporary finish times $\widetilde{end}_{k}^{(i)}$ of each
  worker if it would receive from $sender$ the $i$-th task.
\item
  A worker with the smallest temporary finish time $\widetilde{end}_{k}^{(i)}$ will be the
  receiver, called $receiver$. If there are multiple workers with the same
  temporary finish time $\widetilde{end}_{k}^{(i)}$, we
  take the worker with the smallest finish time $end_{k}^{(i-1)}$.
\item
  If the finish time of $sender$ is strictly larger than the temporary finish time
  $\widetilde{end}_{sender}^{(i)}$ of $sender$, $sender$ sends one task to
  $receiver$ and iterate. Otherwise stop.
\end{itemize}

\begin{lemma}
\label{lemma:receiver}
  On homogeneous star-platforms, in iteration $i$ the \greedyalgo (Algorithm~\ref{algo:star_hom}) always chooses
  as receiver a worker which finishes processing the first in iteration $i-1$.
\end{lemma}

\begin{proof}
  As the platform is homogeneous, all communications take the same time
  and all computations take the same time. In Algorithm~\ref{algo:star_hom} the master
  chooses as receiver in iteration $i$ the worker $k$ that would end the earliest the processing of the
  $i$-th task sent. To prove that worker $k$ is also the worker which finishes
  processing in iteration $i-1$ first, we have to consider two cases:

  \begin{itemize}
  \item\textbf{Task i arrives when all workers are still working.}\\
    As all workers are still working when
    the master finishes to send task $i$, the master chooses as receiver a worker which
    finishes processing the
    first, because this worker will also finish
    processing task $i$ first, as we have homogeneous conditions. See Figure~\ref{fig:lemma_working} for an example: the master chooses worker $k$ as
    in iteration $i-1$ it finishes before worker $j$ and it can thus start
    computing task $i+1$ earlier than worker $j$ could do.

  \item\textbf{Task i arrives when some workers have finished working.}\\
    If some workers have finished working when the master can finish to send task $i$, we are in
    the situation of Figure~\ref{fig:lemma_finished}: All these workers could start
    processing task $i$ at the same time. As our algorithm chooses in this
    case a worker which finished processing first (see line 13 in Algorithm~\ref{algo:star_hom}), the master chooses worker $k$ in the example.
    \qedhere
  \end{itemize}

  \begin{figure}[htbp]
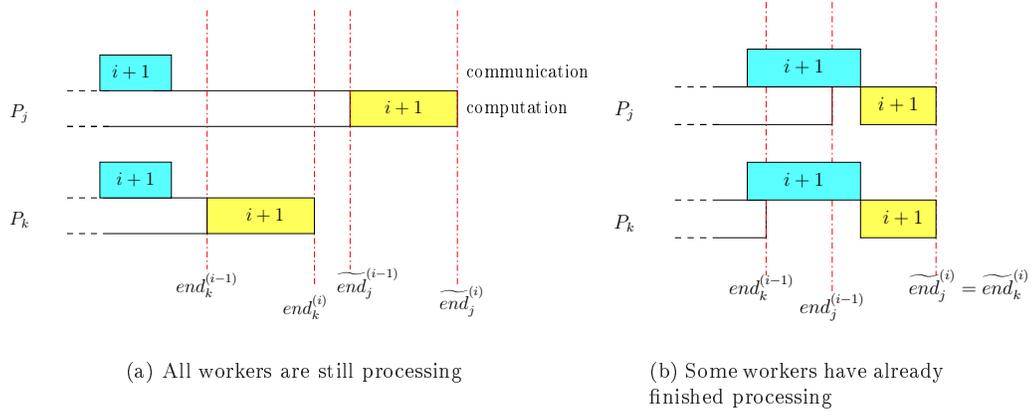

    \centering
    \subfigure[All workers are still processing]{
      \includegraphics[width=0.5\textwidth]{lemmaCleqW.fig}%
      \label{fig:lemma_working}%
    } $\quad$
    \subfigure[Some workers have already finished processing]{%
      \includegraphics[width=0.31\textwidth]{lemmaCgeqW.fig}%
      \label{fig:lemma_finished}%
    }
    \caption{In iteration $i$: The master chooses which worker will be the receiver of
      task $i$.}
    \label{fig:lemma}
  \end{figure}

\end{proof}

The aim of these schedules is always to minimize the makespan. So workers
who take a long time to process their tasks are interested in sending some
tasks to other workers which are less charged in order to decrease their processing
time. If a weakly charged worker sends some tasks to another worker this will
not decrease the global makespan, as a strongly charged worker has still its
long processing time or its processing time might even have increased if it
was the receiver. So it might happen that the weakly charged worker who sent a
task will receive another task in another scheduling step. In the following
lemma we will show that this kind of schedule, where sending workers also
receive tasks, can be transformed in a schedule where this effect does not appear.

\begin{lemma}
\label{lemma:no_recv}
On a platform with homogeneous communications, if there exists a schedule $S$ with makespan $M$, then there also exists a
schedule $S'$ with a makespan $M' \leq M$ such that no worker both sends and
receives tasks.
\end{lemma}

\begin{figure}[htb!]
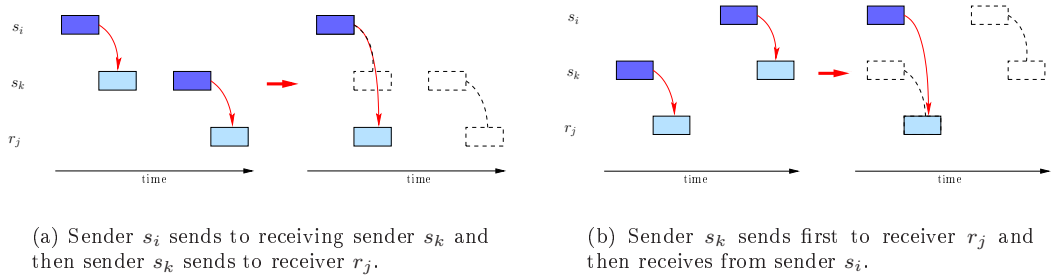

  \centering
  \subfigure[Sender $s_i$ sends to receiving sender $s_k$ and
  then sender $s_k$ sends to receiver $r_j$.]{
    \includegraphics[width=0.44\textwidth]{simpleCase.fig}
    \label{fig:simpleCase}%
  } $\quad$
  \subfigure[Sender $s_k$ sends first to receiver $r_j$ and then receives from
  sender $s_i$.]{
    \includegraphics[width=0.43\textwidth]{difficultCase.fig}
    \label{fig:difficultCase}%
  }
\caption{How to break up sending chains, dark colored communications are
  emissions, light colored communications represent receptions.}
\end{figure}

\begin{proof}
We will prove that we can transform a schedule where senders might receive
tasks in a schedule with equal or smaller makespan where senders do not
receive any tasks.

  \begin{figure}[htbp]
    \centering
    \includegraphics[width=0.45\textwidth]{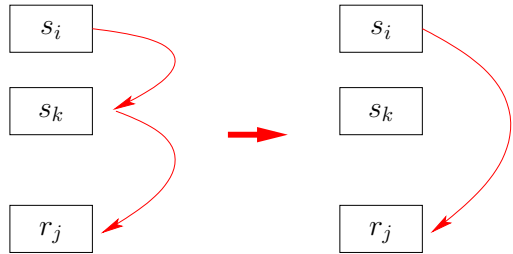}%
    \caption{Scheme on how to break up sending chains.}
    \label{fig:chain}
  \end{figure}

  If the master receives its $i$-th task from processor $P_j$ and sends it to
  processor $P_k$, we say that $P_k$ receives this task from processor $P_j$.

Whatever the schedule, if a sender receives a task we have the situation of a
sending chain (see Figure~\ref{fig:chain}): at some step of the schedule a
sender $s_i$ sends to a sender $s_k$, while in another step of the schedule
the sender $s_k$ sends to a receiver $r_j$. So the master is occupied twice. As all receivers receive in fact their tasks from the
master, it does not make a difference for them which sender sent the task to
the master. So we can break up the sending chain in the following way:
We look for the earliest time, when a sending worker, $s_k$, receives a task from a
sender, $s_i$. Let $r_j$ be a receiver that receives a task from sender $s_k$. There are
two possible situations:
\begin{enumerate}
\item
Sender $s_i$ sends to sender $s_k$ and later sender $s_k$ sends to receiver
$r_j$, see Figure~\ref{fig:simpleCase}. This case is simple: As the communication from
$s_i$ to $s_k$ takes place first and we have homogeneous communication links, we can replace this communication by an emission
from sender $s_i$ to receiver $r_j$ and just delete the second
communication.

\item
Sender $s_k$ sends to receiver $r_j$ and later sender $s_i$
sends to sender
$s_k$, see Figure~\ref{fig:difficultCase}. In this case the reception on receiver $r_j$ happens
earlier than the emission of sender $s_i$, so we can not use exactly the same
mechanism as in the previous case. But we can use our hypothesis that sender
$s_k$ is the first sender that receives a task. Therefore, sender $s_i$ did
not receive any task until $s_k$ receives. So at the moment when $s_k$ sends
to $r_j$, we know that sender $s_i$ already owns the task that it
will send later to sender $s_k$. As we use homogeneous communications, we can
schedule the communication $s_i \rightarrow r_j$ when the communication $s_k
\rightarrow r_j$ originally took place and
delete the sending from $s_i$ to $s_k$.

\end{enumerate}
As in both cases we gain in communication
time, but we keep the same computation time, we do not increase the makespan
of the schedule, but we transformed it in a schedule with one less sending
chain. By repeating this procedure for all sending chains, we transform the
schedule $S$ in a schedule $S'$ without sending chains while not increasing the makespan.
\end{proof}

\begin{proposition}
\greedyalgo (Algorithm~\ref{algo:star_hom}) calculates an optimal schedule $S$ on a homogeneous
star network, where all tasks are initially located on the workers and
communication capabilities as well as computation capabilities are homogeneous and all
tasks have the same size.
\end{proposition}

\begin{algorithm}[htbp]
  \caption{\greedyalgo}
  \label{algo:star_hom}
  \begin{algorithmic}[1]
    \STATE {\small /* initialization */}
    \STATE $i\gets 0$
    \STATE $master\_in^{(i)}\gets 0$
    \STATE $master\_out^{(i)}\gets 0$
    \STATE $\forall{k}\;L_{k}^{(0)} \gets L_{k}$
    \STATE $end_{k}^{(0)} \gets L_{k}^{(0)} \times w$
    \STATE {\small /* the scheduling */}
    \WHILE{true}
    \STATE $sender \gets \max_{k} end_{k}^{(i)}$
    \STATE $master\_in^{(i+1)} \gets master\_in^{(i)} + c$
    \STATE $task\_arrival\_worker = \max (master\_in^{(i+1)}, master\_out^{(i)}) + c$
    \STATE $\forall k$ $\widetilde{end}_{k}^{(i+1)} \gets \max(end_{k}^{(i+1)}, task\_arrival\_worker) + w$
    \STATE select $receiver$ such that $\widetilde{end}_{receiver}^{(i+1)} =
    \min_k \widetilde{end}_{k}^{(i+1)}$ and if there are several processors with the same
    minimum $\widetilde{end}_{k}^{(i+1)}$, choose one with the smallest $end_{k}^{(i)}$
    \IF{$end_{sender}^{(i)} \leq \widetilde{end}_{receiver}^{(k+1)}$}
    \STATE {\small /* we can not improve the makespan anymore */}
    \STATE break
    \ELSE
    \STATE {\small /* we improve the makespan by sending the task to the $receiver$ */}
    \STATE $master\_out^{(i+1)}\gets task\_arrival\_worker$
    \STATE $end_{sender}^{(i+1)} \gets end_{sender}^{(i)} - w$
    \STATE $L_{sender}^{(i+1)} \gets L_{sender}^{(i)} - 1$
    \STATE $end_{receiver}^{(i+1)} \gets \widetilde{end}_{receiver}^{(i+1)}$
    \STATE $L_{receiver}^{(i+1)} \gets L_{receiver}^{(i)} + 1$
    \FORALL{$j \neq receiver$ and $j \neq sender$}
    \STATE $end_{j}^{(i+1)} \gets end_{j}^{(i)}$
    \STATE $L_{j}^{(i+1)} \gets L_{j}^{(i)}$
    \ENDFOR
    \STATE $i \gets i+1$
    \ENDIF
    \ENDWHILE

  \end{algorithmic}
\end{algorithm}

\begin{proof}
  To prove that \greedy is optimal, we take a schedule $S_{algo}$
  calculated by Algorithm~\ref{algo:star_hom}. Then we take an optimal
  schedule $S_{opt}$. (Because of Lemma~\ref{lemma:no_recv} we can assume that in the schedule $S_{opt}$ no worker
  both sends and receives tasks.) We are going to transform  by induction this optimal schedule into our
  schedule $S_{algo}$.

  As we use a homogeneous platform, all workers have the same
  communication time $c$. Without loss of generality, we can assume that both
  algorithms do all communications as soon as possible (see
  Figure~\ref{fig:ASAP}). So we can divide our schedule $S_{algo}$ in $s_{a}$
  steps and $S_{opt}$ in $s_{o}$ steps. A step
  corresponds to the emission of one task, and we number in this order the
  tasks sent. Accordingly the $s$-th task is the task sent during step $s$ and
  the actual schedule corresponds to the load distribution after the $s$ first
  tasks. We start
  our schedule at time $T = 0$.

  \begin{figure}[htb]
    \centering
    \includegraphics[width=0.6\textwidth]{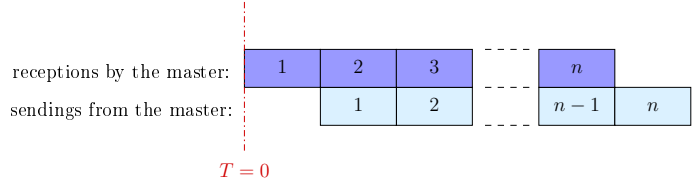}%
    \caption{Occupation of the master.}
    \label{fig:ASAP}
  \end{figure}

  Let $S(i)$ denote the worker receiving the $i$-th task under schedule $S$. Let $i_{0}$ be
  the first step where $S_{opt}$ differs from $S_{algo}$,
  i.e., $S_{algo}(i_{0}) \neq S_{opt}(i_{0})$ and $\forall i < i_{0}$,
  $S_{algo}(i) = S_{opt}(i)$. We look for a step $j > i_{0}$, if it exists, such that
  $S_{opt}(j) = S_{algo}(i_{0})$ and $j$ is minimal.

We are in the following situation: schedule $S_{opt}$ and schedule
$S_{algo}$ are the same for all tasks $[1..(i_0 -1)]$.
As worker $S_{algo}(i_0)$ is chosen at step $i_0$, then, by definition of
Algorithm~\ref{algo:star_hom}, this means that this worker finishes first its processing after the reception of the
$(i_0 -1)$-th tasks (cf. Lemma~\ref{lemma:receiver}). As $S_{opt}$
and $S_{algo}$ differ in step $i_0$, we know that $S_{opt}$ chooses worker $S_{opt}(i_0)$
 that finishes the schedule of its load after step
$(i_0 -1)$ no sooner than worker $S_{algo}(i_0)$.\\

\textbf{Case 1:} Let us first consider the case where there exists such a step $j$.
So $S_{algo}(i_0) =
S_{opt}(j)$ and $j>i_0$. We know that worker $S_{opt}(j)$ under schedule $S_{opt}$ does not receive any task
between step $i_0$ and step $j$ as $j$ is chosen minimal.

  We use the following notations for the schedule $S_{opt}$, depicted on
  Figures~\ref{fig:small_tj},~\ref{fig:big_tj}, and~\ref{fig:string}:
  \begin{description}
  \item[$\mathbf{T_{j}}$:]
    the date at which the reception of task $j$ is finished on worker
    $S_{opt}(j)$, i.e., $T_{j} = j \times c + c$ (the time it takes the master
    to receive the first task plus the time it takes him to send $j$ tasks).
  \item[$\mathbf{T_{i_{0}}}$:]
    the date at which the reception of task $i_{0}$ is finished on worker
    $S_{opt}(i_{0})$, i.e., $T_{i_{0}} = i_{0} \times c + c$.
  \item[$\mathbf{F_{pred(j)}}$:]
    time when computation of task $pred(j)$ is finished, where
    task $pred(j)$ denotes the last task which is computed on worker
    $S_{opt}(j)$ before task $j$ is computed.
  \item[$\mathbf{F_{pred(i_{0})}}$:]
    time when computation of task $pred(i_{0})$ is
    finished, where
    task $pred(i_{0})$ denotes the last task which is computed on worker
    $S_{opt}(i_{0})$ before task $i_{0}$ is computed.
  \end{description}

    We have to consider two sub-cases:
    \begin{itemize}

    \item
      $\mathbf{T_{j} \leq F_{pred(i_{0})}}$ (Figure~\ref{fig:small_before}).\\
This means that we are in the following situation: the reception of task $j$
      on worker  $S_{opt}(j)$  has already finished when worker $S_{opt}(i_0)$ finishes
      the work it has been scheduled until step $i_0-1$.

In this case we exchange the tasks $i_0$ and $j$ of schedule
$S_{opt}$ and we create the following schedule $S_{opt}'$:\\
      $S_{opt}'(i_{0}) = S_{opt}(j) = S_{algo}(i_0)$,\\
      $S_{opt}'(j) = S_{opt}(i_{0})$
      \\
      and $\forall i\neq i_{0},j,\; S_{opt}'(i) = S_{opt}(i)$. The schedule of
      the other workers is kept unchanged.
      All tasks are executed at the same date than previously (but maybe not
      on the same processor).

      \begin{figure}[htbp]
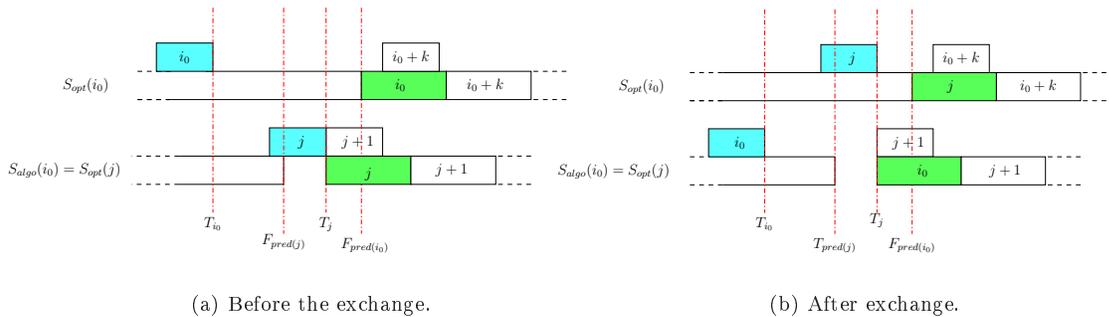

        \centering
        \subfigure[Before the exchange.]{
          \includegraphics[width=0.45\textwidth]{smallTj.fig}%
          \label{fig:small_before}%
        } $\quad$
        \subfigure[After exchange.]{%
          \includegraphics[width=0.45\textwidth]{smallTjAfter.fig}%
          \label{fig:small_after}%
        }
        \caption{Schedule $S_{opt}$ before and after exchange of tasks
          $i_{0}$ and $j$.}
      \label{fig:small_tj}
    \end{figure}

      Now we prove that this kind of exchange is possible.

      We
      know that worker $S_{opt}(j)$ is not scheduled any task later than step
      $i_0 -1$ and before step $j$, by definition of $j$. So we know that this worker can start
      processing task $j$ when task $j$ has arrived and when it has
      finished processing its amount of work scheduled until step
      $i_0-1$.
      We already know
      that worker $S_{opt}(j) = S_{algo}(i_0)$ finishes processing its tasks scheduled until
      step $i_0-1$ at a time earlier than or equal to that of worker
      $S_{opt}(i_0)$ (cf. Lemma~\ref{lemma:receiver}). As we are in
      homogeneous conditions, communications and processing of a task takes
      the same time on all processors. So we can exchange the destinations of
      steps $i_0$ and $j$ and keep the same moments of execution, as both tasks will arrive in time to be processed
      on the other worker: task $i_0$ will arrive at worker $S_{opt}(j)$ when
      it is still processing and the same for task $j$ on worker $S_{opt}(i_0)$. Hence task $i_0$ will be sent to worker
      $S_{opt}(j) = S_{algo}(i_0)$ and
      worker $S_{opt}(i_0)$ will receive task $j$.
      So schedule $S_{opt}$ and schedule $S_{algo}$ are the same for all tasks $[1
      .. i_0]$ now.
      As both tasks arrive in time and can be executed instead of the other task, we
      do not change anything in the makespan $M$. And as $S_{opt}$ is optimal, we
      keep the optimal makespan.

    \item
      $\mathbf{T_{j} \geq F_{pred(i_{0})}}$ (Figure~\ref{fig:before}).\\
      In this case we have the following situation: task $j$ arrives on worker
      $S_{opt}(j)$, when worker  $S_{opt}(i_0)$ has already finished
      processing its tasks scheduled until step $i_0 -1$.\\
      In this case we exchange the schedule destinations $i_{0}$ and $j$ of schedule $S_{opt}$
      beginning at tasks $i_{0}$ and $j$ (see Figure~\ref{fig:big_tj}).
      In other words we create a schedule $S_{opt}'$:\\
      $\forall i\geq i_{0}$ such that $S_{opt}(i) = S_{opt}(i_{0})$:
      $S_{opt}'(i) = S_{opt}(j) = S_{algo}(i_0)$
      \\
      $\forall i\geq j$ such that $S_{opt}(i) = S_{opt}(j)$:
      $S_{opt}'(i) = S_{opt}(i_{0})$
      \\
      and $\forall i\leq i_{0}$ $S_{opt}'(i) = S_{opt}(i)$. The schedule $S_{opt}$ of
      the other workers is kept unchanged. We recompute the
      finish times $F_{S_{opt}}^{(s)}(j)$ of workers $S_{opt}(j)$ and $S_{opt}(i_0)$ for all steps $s
      > i_{0}$.

      \begin{figure}[htbp]
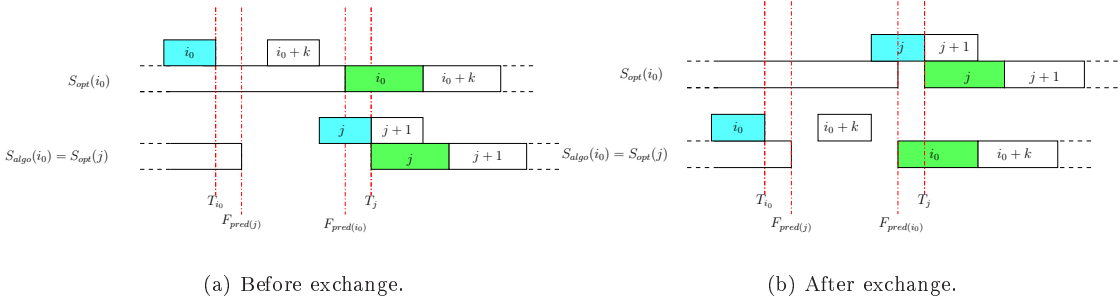

        \centering
        \subfigure[Before exchange.]{
          \includegraphics[width=0.45\textwidth]{big_tj.fig}%
          \label{fig:before}%
        } $\quad$
        \subfigure[After exchange.]{
          \includegraphics[width=0.45\textwidth]{big_tj_after.fig}%
          \label{fig:after}%
        }
        \caption{Schedule $S_{opt}$ before and after exchange of lines
          $i_{0}$ and $j$.}
        \label{fig:big_tj}
      \end{figure}

      Now we prove that this kind of exchange is possible. First of all we know that worker $S_{algo}(i_0)$ is the same as the worker
      chosen in step $j$ under schedule $S_{opt}$ and so $S_{algo}(i_0) =
      S_{opt}(j)$. We also know that worker $S_{opt}(j)$ is not scheduled any tasks later than step
      $i_0 -1$ and before step $j$, by definition of $j$. Because of the
      choice of worker $S_{algo}(i_0) = S_{opt}(j)$ in $S_{algo}$, we know that worker
      $S_{opt}(j)$ has finished working when task $j$ arrives: at step $i_0$
      worker $S_{opt}(j)$ finishes earlier than or at the same time as worker
      $S_{opt}(i_0)$ (Lemma~\ref{lemma:receiver}) and as we are in
      the case where $T_{j} \geq F_{pred(i_{0})}$, $S_{opt}(j)$ has also
      finished when $j$ arrives. So we can exchange the
      destinations of the workers  $S_{opt}(i_0)$ and $S_{opt}(j)$ in the
      schedule steps equal to, or later than, step $i_0$ and process them at
      the same time as we would do on the other worker.
      As we have shown that we can start processing task $j$ on worker
      $S_{opt}(i_0)$ at the same time as we did on worker $S_{opt}(j)$, and
      the same for task $i_0$, we keep the same makespan.  And as $S_{opt}$ is optimal, we
      keep the optimal makespan.

\end{itemize}
   \textbf{Case 2:} If there does not exist a $j$, i.e., we can not find a
    schedule step $j>i_0$ such that worker $S_{algo}(i_0)$ is scheduled a task
    under schedule $S_{opt}$, so we know that no other task
    will be scheduled on worker $S_{algo}(i_0)$ under the schedule $S_{opt}$. As our algorithm chooses in step $s$ the worker
    that finishes task $s+1$ the first, we know that worker  $S_{algo}(i_0)$
    finishes at a time earlier or equal to that of $S_{opt}$. Worker $S_{algo}(i_0)$ will be
    idle in the schedule $S_{opt}$ for the
    rest of the algorithm, because otherwise we would have found a step $j$. As we
    are in homogeneous conditions, we can simply
    displace task $i_{0}$ from worker $S_{opt}(i_{0})$ to worker
    $S_{algo}(i_{0})$ (see Figure~\ref{fig:string}). As we have
    $S_{opt}(i_{0}) \neq S_{algo}(i_{0})$ and with Lemma~\ref{lemma:receiver}
    we know that worker $S_{algo}(i_{0})$ finishes processing its tasks until
    step $i_0 -1$ at a time earlier than or equal to $S_{opt}(i_{0})$, and we do not downgrade
    the execution time because we are in homogeneous conditions.\\

      \begin{figure}[htbp]
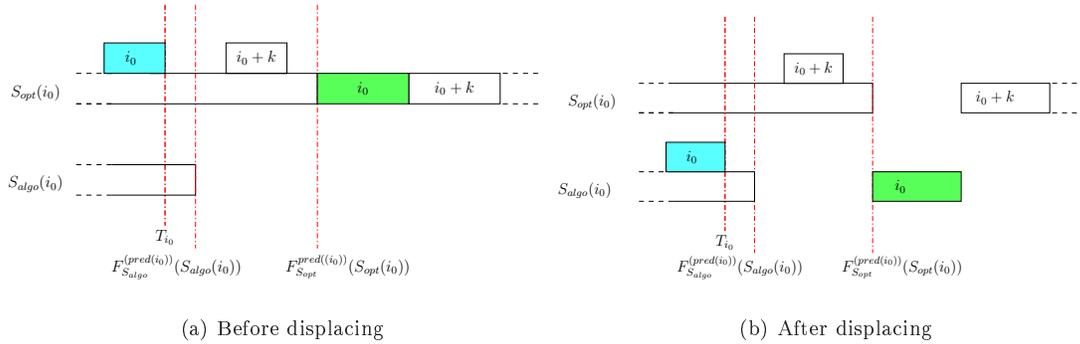

        \centering
        \subfigure[Before displacing]{
          \includegraphics[width=0.45\textwidth]{noJ.fig}%
          \label{fig:string_before}%
        } $\quad$
        \subfigure[After displacing]{%
          \includegraphics[width=0.45\textwidth]{noJAfter.fig}%
          \label{fig:string_after}%
        }
        \caption{Schedule $S_{opt}$ before and after displacing task $i_{0}$.}
        \label{fig:string}
      \end{figure}

    Once we have done the exchange of task $i_{0}$, the schedules $S_{opt}$ and
    $S_{algo}$ are the same for all tasks $[1 .. i_{0}]$. We restart the
    transformation until $S_{opt} = S_{algo}$ for all tasks $[1 .. \min(s_{a},s_o)]$
    scheduled by $S_{algo}$.\\

Now we will prove by contradiction that the number of tasks scheduled by
$S_{algo}$, $s_{a}$, and $S_{opt}$, $s_o$, are the same. After $\min(s_{a},s_o)$ transformation
steps  $S_{opt} = S_{algo}$ for all tasks $[1 .. \min(s_{a},s_o)]$
scheduled by $S_{algo}$.
So if after these steps  $S_{opt} = S_{algo}$ for all $n$
tasks, both algorithms redistributed the same number of tasks and we
have finished.

We now consider the case $s_a \neq s_o$.
In the case of $s_a > s_o$, $S_{algo}$ schedules more tasks than
$S_{opt}$. At each step of our algorithm we do not increase the makespan. So if
we do more steps than $S_{opt}$, this means that we scheduled some tasks
without changing the global makespan. Hence $S_{algo}$ is optimal.

If  $s_a < s_o$, this means that $S_{opt}$ schedules more tasks than
$S_{algo}$ does. In this case,
after $s_a$ transformation steps, $S_{opt}$ still schedules tasks.
If we take a look at the schedule of the $(s_a +1) $-th task in
$S_{opt}$: regardless which receiver $S_{opt}$ chooses, it will increase the
makespan as we prove now. In the
following we will call $s_{algo}$ the worker our algorithm would have chosen
to be the sender, $r_{algo}$ the worker our algorithm would have chosen to be
the receiver. $s_{opt}$
and $r_{opt}$ are the sender and receiver chosen by the optimal schedule. Indeed, in our algorithm we would have chosen $s_{algo}$ as sender such
that it is a worker which finishes last. So the time worker $s_{algo}$
finishes processing is $F_{s_{algo}} =
M(S_{algo})$. $S_{algo}$ chooses the receiver $r_{algo}$ such that it finishes
processing the received
task the earliest of all possible receivers and such that it also finishes processing the
receiving task at the same time or earlier than the sender would do. As
$S_{algo}$ did not decide to send the $(s_a +1) $-th task, this
means, that it could not find a receiver which fitted. Hence we
know, regardless which receiver $S_{opt}$ chooses, that the makespan will strictly
increase (as $S_{algo} = S_{opt}$ for all $[1 .. s_a]$).
We take a look at the makespan of $S_{algo}$ if we would have
scheduled the  $(s_a +1) $-th task. We know that we can not
decrease the makespan anymore, because in our algorithm we decided to keep the
schedule unchanged. So after
the emission of the $(s_a +1) $-th task, the makespan would become $M(S_{algo})
= F_{r_{algo}} \geq F_{s_{algo}}$. And $F_{r_{algo}}\leq F_{r_{opt}}$,
because of the definition of receiver ${r_{algo}}$. As $M(s_{opt}) \geq F_{r_{opt}}$,
we have $M(S_{algo}) \leq
M(S_{opt})$. But we decided not to do this schedule as $M(S_{algo})$ is smaller
before the schedule
of the $(s_a +1) $-th task than afterwards. Hence we
get that $M(S_{algo}) < M(S_{opt})$. So the only possibility why $S_{opt}$
sends the $(s_a +1) $-th task and still be optimal is that, later on,
$r_{opt}$ sends a task to some other processor $r_k$. (Note that even
  if we choose $S_{opt}$ to have no such chains in the beginning, some might
  have appeared because of our previous transformations). In the same manner as we transformed sending
chains in Lemma~\ref{lemma:no_recv}, we can suppress this sending chain, by
sending task $(s_a +1)$ directly to $r_{k}$ instead of sending to $r_{opt}$. With the same
argumentation, we do this by induction for all tasks $k$, $(s_a + 1) \leq k \leq s_o$, until schedule $S_{opt}$
and $S_{algo}$ have the same number $s_o = s_a$ and so $S_{opt} = S_{algo}$
and hence $M(S_{opt}) = M(S_{algo})$.
\end{proof}

\paragraph{Complexity:}
The initialization phase is in $O(m)$, as we have to compute the finish times
for each worker. The while loop can be run at maximum $n$ times, as we can
not redistribute more than the $n$ tasks of the system. Each iteration is in the
order of $O(m)$, which leads us to a total run time of $O(m\times n)$.

\subsection{Scheduling on platforms with homogeneous
  communication links and heterogeneous computation capacities}
\label{sec:moore}
In this section we present an algorithm for star-platforms with homogeneous
communications and heterogeneous workers, the \moorealgo (\moore). For a given makespan, we compute if
there exists a possible schedule to finish all work in time. If there is one,
we optimize the makespan by a binary search. The plan of the section is as
follows: In Section~\ref{sec:moore_original} we present an existing algorithm
which will be the basis of our work. The framework and some usefull notations
are introduced in Section~\ref{sec:notations_moore}, whereas the real
algorithm is the subject of Section~\ref{sec:star_plat}.

\subsubsection{Moore's algorithm}
\label{sec:moore_original}
In this section we present \textsc{Moore's algorithm} \cite{Brucker:2004, Moore:1968}, whose aim is to maximize the number of tasks to be processed
in-time, i.e., before tasks exceed their deadlines. This algorithm gives a solution
to the $1||\sum U_j$ problem when the maximum number, among $n$ tasks, has
to be processed in time
on a single machine. Each task $k$, $1\leq k \leq n$, has a processing time
$w_k$ and a deadline $d_{k}$, before which it has to be
processed.

\paragraph{Moore's algorithm works as follows:}
All tasks are ordered in non-decreasing order of their deadlines. Tasks are
added to the solution one by one in this order as long as their deadlines are
satisfied. If a task $k$ is out of time, the task $j$ in the actual solution with the largest
processing time $w_j$ is deleted from the solution.

Algorithm~\ref{algo:exist} \cite{Brucker:2004, Moore:1968} solves in $O(n \log n)$ the $1||\sum U_j$ problem:
it constructs a maximal set $\sigma$ of early jobs.

\begin{algorithm}[htbp]
  \caption{Moore's algorithm}
  \label{algo:exist}
  \begin{algorithmic}[1]
    \STATE Order the jobs by non-decreasing deadlines: $d_1 \leq d_2 \leq \dots
    \leq d_d$
    \STATE $\sigma \gets \emptyset$; $t \gets 0$
    \FOR{$i:=1$ to $n$}
    \STATE $\sigma \gets \sigma \cup \{i\}$
    \STATE $t \gets t + w_i$
    \IF{$t > d_i$}
    \STATE Find job $j$ in $\sigma$ with largest $w_j$ value
    \STATE $\sigma \gets \sigma\backslash \{j\}$
    \STATE $t \gets t-w_j$
    \ENDIF
    \ENDFOR
  \end{algorithmic}
\end{algorithm}

\subsubsection{Framework and notations for \moore}
\label{sec:notations_moore}
We keep the star network of Section~\ref{sec:framework_tasks} with homogeneous
communication links. In contrast to Section~\ref{sec:greedy} we suppose
$m$ heterogeneous workers who own initially a number $L_i$ of identical
independent tasks.

Let $M$ denote the objective makespan for the searched schedule $\sigma$ and
  $f_{i}$ the time needed by worker $i$ to process its initial load. During
  the algorithm execution we divide all workers in two subsets, where $S$ is the set of
  senders ($s_{i} \in S$ if $f_{i} > M$) and $R$ the set of receivers ($r_{i} \in R$ if $f_{i} < M$).
As our algorithm is based on Moore's, we need a notation for deadlines. Let
  $d_{r_{i}}^{(k)}$ be the deadline to receive the $k$-th task on receiver $r_i$.
  $l_{s_{i}}$ denotes the number of tasks sender $i$ sends to the master and
  $l_{r_{i}}$ stores the number of tasks receiver $i$ is able to receive from
  the master. With help of these values we can determine the total amount of
  tasks that must be sent as
  $L_{send} = \sum_{s_{i}}l_{s_{i}}$. The total amount of task if all receivers
  receive the maximum amount of tasks they are able to receive is
  $L_{recv} = \sum_{r_{i}}l_{r_{i}}$. Finally, let $L_{sched}$ be the maximal amount of tasks that can be scheduled by the algorithm.

\subsubsection{Moore based binary search algorithm - \moore}
\label{sec:star_plat}

\paragraph{Principle of the algorithm:}
Considering the given makespan we determine overcharged workers, which can not
finish all their tasks within this makespan. These overcharged workers will then
send some tasks to undercharged workers, such that all of them can finish
processing within the makespan. The algorithm solves the following two
questions: Is there a possible schedule such that all workers can finish in
the given makespan? In which order do we have to send and receive to obtain
such a schedule?

\paragraph{The algorithm can be divided into four phases:}

\begin{description}
\item[Phase 1] decides which of the workers will be senders and which receivers, depending
  of the given makespan (see Figure~\ref{fig:initialDist}). Senders are workers which are not able to process all
  their initial tasks in time, whereas receivers are workers which could treat more tasks in
  the given makespan $M$ than they hold initially. So sender $P_i$ has a
  finish time $f_i > M$, i.e., the time needed to compute their initial
  tasks is larger than the given makespan $M$. Conversely, $P_i$ is a receiver
  if it has
  a finish time $f_i < M$. So the set of senders in the example
  of Figure~\ref{fig:initialDist} contains $s_1$ and $s_v$, and the set of receivers $r_1, r_2$, and $r_u$.

  \begin{figure}[htb!]
    \centering
    \includegraphics[width=0.7\textwidth]{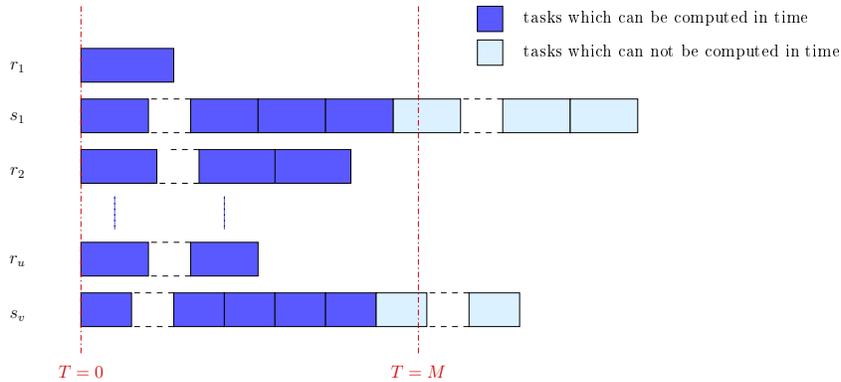}
    \caption{Initial distribution of the tasks to the workers, dark colored tasks
      can be computed in-time, light colored tasks will be late and have to be
      scheduled on some other workers.}
    \label{fig:initialDist}
  \end{figure}

\item[Phase 2] fixes how many transfers have to be scheduled from each sender
  such that the senders all finish their remaining tasks in time. Sender $s_i$ will
  have to send an amount of tasks $l_{s_{i}} = \left\lceil \frac{f_{s_{i}} -
    T}{w_{s_{i}}} \right\rceil$ (i.e., the number of light colored tasks of a sender
  in Figure~\ref{fig:initialDist}).

\item[Phase 3] computes for each receiver the deadline of each of the tasks it can receive, i.e., a pair $(d_{r_{j}}^{(i)},
  r_j)$ that
  denotes the $i$-th deadline of receiver $r_{j}$. Beginning at the makespan $M$ one
measures when the last task has to arrive on the receiver such that it can be processed
in time. So the latest moment that a task can arrive so that it can still be computed
on receiver $r_j$ is $T-w_{r_{j}}$, and so on. See Figure
\ref{fig:deadlines} for an example.

\begin{figure}[htb!]
  \centering
  \includegraphics[width=0.6\textwidth]{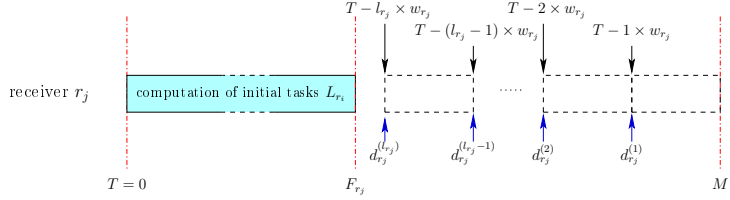}
  \caption{Computation of the deadlines $d_{r_{j}}^{(k)}$ for worker $r_{j}$.}
  \label{fig:deadlines}
\end{figure}

\item[Phase 4] is the proper scheduling step: The master decides which tasks have to
be scheduled on which receivers and in which order. In this phase we use
Moore's algorithm. Starting at time $t = c$ (this is the time,
when the first task arrives at the master), the master can start scheduling
the tasks on the receivers. For this purpose the deadlines $(d, r_{j})$
are ordered by non-decreasing $d$-values. In the same manner as in Moore's
algorithm, an optimal schedule $\sigma$ is computed by adding one by one tasks
to the schedule: if we consider the deadline $(d, r_j)$, we add a task to
processor $r_j$. The
corresponding processing time is the communication time $c$ of
$r_{j}$. So if a deadline is not met, the last reception is suppressed from
$\sigma$ and we continue. If the schedule is able to send at least $L_{send}$
tasks the algorithm succeeds, otherwise it fails.

  \end{description}
Algorithm~\ref{algo:star_het} describes \moore in pseudo-code. Note that the algorithm
is written for heterogeneous conditions, but here we study it for
homogeneous communication links.

\begin{algorithm}[htbp]
  \caption{Algorithm for star-platforms with homogeneous communications and heterogeneous workers}
  \label{algo:star_het}

  \begin{algorithmic}[1]
    \STATE /* Phase 1: Initialization */
    \STATE  initialize $f_{i}$ for all workers $i$, $f_{i} = L_{i} \times
    w_{i}$
    \STATE compute $R$ and $S$, order $S$ by non-decreasing values $c_{i}$ such
    that $c_{s_{1}} \leq c_{s_{2}} \leq \dots$

    \STATE /* Phase 2: Preparing the senders */
    \FORALL{$s_{i} \in S$}
    \STATE $l_{s_{i}} \gets \left\lceil \frac{f_{s_{i}} -
      T}{w_{s_{i}}} \right\rceil$
    \IF{$\left\lfloor \frac{T}{c_{s_{i}}}\right\rfloor < l_{s_{i}}$}
    \STATE /* $M$ too small */
    \STATE return ($false, \emptyset$)
    \ENDIF
    \ENDFOR
    \STATE total number of tasks to send: $L_{send} \gets \sum_{s_{i}}l_{s_{i}}$

    \STATE /* Phase 3: Preparing the receivers */
    \STATE $D \gets \emptyset$
    \FORALL{$r_{i} \in R$}
    \STATE $l_{r_{i}}\gets 0$
    \WHILE{$f_{r_{i}} \leq M - (l_{r_{i}} + 1) \times w_{r_{i}}$}
    \STATE $l_{r_{i}} \gets l_{r_{i}} +1$
    \STATE $d_{r_{i}}^{(l_{r_{i}})} \gets M - (l_{r_{i}}\times w_{r_{i}})$
    \STATE $D \gets D \cup {(d_{r_{i}}^{(l_{r_{i}})}, r_{i})}$
    \ENDWHILE
    \ENDFOR
    \STATE number of tasks that can be received: $L_{recv} \gets \sum_{r_{i}}l_{r_{i}}$

    \STATE /* Phase 4: The master schedules */
    \STATE senders send in non-decreasing order of values $c_{s_{i}}$ to the
    master
    \STATE order deadline-list $D$ by non-decreasing values of
    deadlines $d_{r_i}$ and rename the deadlines in this order from $1$ to $L_{recv}$
    \STATE $\sigma \gets \emptyset$; $t\gets c_{s_1}$; $L_{sched} = 0$;
    \FOR{$i = 1$ to $L_{recv}$}
    \STATE $(d_{i},r_{i}) \gets i$-th element $(d_{r_{k}}^{(j)}, r_{k})$ of $D$
    \STATE $\sigma \gets \sigma \cup \{r_{i}\}$
    \STATE $t \gets t + c_{r_{i}}$
    \STATE $L_{sched} \gets L_{sched}+1$
    \IF{$t > d_{i}$}
    \STATE Find $(d_j, r_{j})$ in $\sigma$ such that $c_{r_j}$ value is largest
    \STATE $\sigma \gets \sigma \backslash \{(d_j, r_j)\}$
    \STATE $t \gets t - c_{r_j}$
    \STATE $L_{sched} \gets L_{sched} -1$
    \ENDIF
    \ENDFOR
    \STATE return $((L_{sched} \geq L_{send}), \sigma)$
  \end{algorithmic}
\end{algorithm}

 \begin{theorem} \moore (Algorithm~\ref{algo:star_het}) succeeds to build a schedule $\sigma$ for a given makespan
   $M$, if and only if there exists a schedule with makespan less than or
   equal to $M$, when the platform is made of one master, several workers with
   heterogeneous computation power but homogeneous communication capabilities.
 \end{theorem}

 \begin{proof}
   Algorithm~\ref{algo:exist} (Moore's Algorithm) constructs a maximal set $\sigma$ of early jobs on a single machine scheduling problem. So we are going to show that
   our algorithm can be reduced to this problem.

   As we work with a platform with homogeneous communications, we do not have to care about the arrival times of jobs
   at the master, apart from the first job.
   Our deadlines correspond to the latest moments, at which tasks can arrive on the
   workers such that they can be processed in-time (see Figure~\ref{fig:deadlines}).
   So we have a certain number $L_{recv}$ of possible receptions for all
   receivers.

Phases 1 to 3 prepare our scheduling problem to be similar to the situation in
Algorithm~\ref{algo:exist} and thus to be able to use it.

In phase 1 we distinguish which processors have to be senders, which have to
be receivers. With Lemma~\ref{lemma:no_recv} we know that we can partition
our workers in senders and receivers (and workers which are none of both), because senders will never receive any
tasks. Phase 2 computes the number of tasks $L_{send}$ that has to be
scheduled. Phase 3 computes the $(d_{r_{j}}^{(k)}, r_{j})$-values, i.e., the
deadlines $d_{r_{j}}^{(k)}$ for each receiver $r_{j}$.
Step 4 is the proper scheduling step and it
corresponds to Moore's algorithm. It computes a maximal set
$\sigma$ of in-time jobs, where $L_{sched}$ is the number of scheduled tasks.

The algorithm returns \textbf{true} if the number of scheduled tasks
$L_{sched}$ is bigger than, or equal to, the number of tasks to be sent $L_{send}$.

Now we will prove that if
there exists a schedule whose makespan is less than, or equal to, $M$,
Algorithm~\ref{algo:star_het} builds one and returns
\textbf{true}.
Consider an optimal schedule $\sigma^*$ with a makespan $M$. We will prove
that Algorithm~\ref{algo:star_het} will return \textbf{true}.

We have computed, for each receiver $r_{j}$, $l_{r_{j}}$ the maximal number of tasks $r_j$ can
process after having finished to process its initial load. Let $N_{r_{j}}$
denote the number of tasks received by $r_{j}$ in $\sigma^*$, $N_{r_j} \leq l_{r_j}$.
For all receivers $r_{j}$ we
know the number $N_{r_j}$ of scheduled tasks. So we have $L^*_{sched} =
\sum_{r_{j}}N_{r_{j}}$. As in an optimal schedule all
tasks sent by the senders are processed on the receivers, we know that
$L^*_{sched} = L_{send}^*$. Let us denote $D$ the set of deadlines computed in our algorithm for the
scheduling problem of which $\sigma^*$ is an optimal solution. We also define
the following set
$D^* = \bigcup_i \bigcup_{1\leq j \leq  N_{r_i}}(M - j\times w_{r_i},r_i)$ of the $N_{r_j}$ latest
  deadlines for each receiver $r_j$. Obviously $D* \subseteq D$. The set of tasks in $\sigma^*$ is exactly a set of tasks
  that respects the deadlines in $D^*$.
The application of the algorithm of Moore on the same problem returns
a maximal solution if there exists a solution. With  $D^* \subset D$,
we already know that there exists a solution with $L_{sched}^*$
scheduled tasks. So Moore's algorithm will return a solution with $L_{sched} \geq L_{sched}*$,
as there are more possible deadlines. On the other side, we have $L_{send}^* \geq
L_{send}$ as $L_{send}$ is the minimal number of tasks that have to be sent
to fit in the given makespan. So we get that $L_{sched} \geq L_{send}$.
As we return \textbf{true} in our algorithm if $L_{sched} \geq L_{send}$, we
will return \textbf{true} whenever there exists a schedule whose makespan is less than, or equal to, $M$.

\begin{figure}[htb!]
  \centering
  \includegraphics[width=0.6\textwidth]{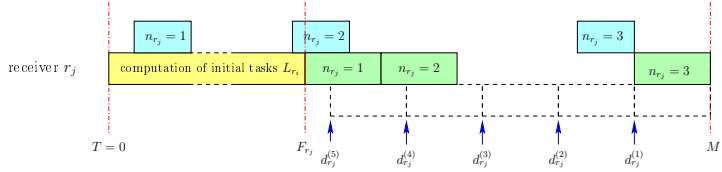}
  \caption{Number of loads scheduled to receiver $r_j$ in order to its deadlines.}
  \label{fig:nTasks}
\end{figure}

Now we prove that if Algorithm~\ref{algo:star_het} returns \textbf{true}
there exists a schedule whose makespan is less than, or equal to, $M$. Our algorithm returns \textbf{true}, if it has found a schedule
$\sigma$ where $L_{sched} \geq L_{send}$. If $L_{sched} = L_{send}$ then the
schedule $\sigma$ found by our algorithm
is a schedule whose makespan is less than, or equal to, $M$. If $L_{sched} >
L_{send}$, we take the $L_{send}$ first
elements of $\sigma$, which still defines a schedule whose makespan is less
than, or equal to, $M$.
\end{proof}

\begin{proposition}
  Algorithm~\ref{algo:opt_T} returns in polynomial time an optimal schedule $\sigma$ for the
  following scheduling problem: minimizing the makespan on a star-platform with homogeneous communication links and heterogeneous
  workers where the initial tasks are located on the workers.
\end{proposition}

\begin{algorithm}[h!]
  \caption{Algorithm to optimize the makespan.}
  \label{algo:opt_T}
  \begin{algorithmic}
    \STATE $/*$ idea: make a binary search of $M \in
    [\min(f_i),\max(f_i)]$ $ */$
    \STATE input: $w_i = \frac{\alpha_i}{\beta_i},\alpha_i,\beta_i \in
    \mathbb{N}\times \mathbb{N}^*$, $c_i = \frac{\gamma_i}{\delta_i},\gamma_i,\delta_i \in
    \mathbb{N}\times \mathbb{N}^*$
    \STATE $\lambda \gets \textnormal{lcm}\{\beta_i,\delta_i\},\; 1\leq i \leq m$
    \STATE $precision \gets \frac{1}{\lambda}$
    \STATE $lo \gets \min(f_i)$; $hi \gets \max(f_i)$;
    \STATE \textbf{procedure {\it binary-Search(lo, hi):}}
    \STATE $gap \gets |lo - hi|$
    \WHILE{$gap > precision$}
    \STATE $M \gets (lo+hi)/2$
    \STATE $found \gets$ \moore($M$)
    \IF{$\not found$}
    \STATE /* $M$ is too small */
    \STATE $lo \gets M$
    \ELSE
    \STATE /* $M$ is maybe too big */
    \STATE $hi \gets M$
    \STATE $\sigma \gets$ found schedule
    \ENDIF
    \STATE $gap \gets |lo - hi|$
    \ENDWHILE
    \STATE return $\sigma$
  \end{algorithmic}
\end{algorithm}

\begin{proof}
We perform a binary search for a solution in a starting interval of
$[\min(f_i),\max(f_i)]$. As we are in heterogeneous computation conditions, we have
heterogeneous $w_i$-values, $1 \leq i \leq m$, $w_i \in \mathbb{Q}$. The
communications instead are homogeneous, so we have $c_i = c$, $1 \leq i \leq m$, $c \in \mathbb{Q}$. Let the
representation of the values be of the following form:
$$w_i = \frac{\alpha_i}{\beta_i}, \alpha_i, \beta_i \in \mathbb{N} \times \mathbb{N}^*,$$
where $\alpha_i$ and $\beta_i$ are prime between each other,
$$c_i = c = \frac{\gamma}{\delta}, \gamma, \delta \in \mathbb{N} \times \mathbb{N}^*,$$
where $\gamma$ and $\delta$ are prime between each other.

Let $\lambda$ be the least common multiple of the denominators $\beta_i$ and
  $\delta_i$, $\lambda = \textnormal{lcm}\{\beta_i, \delta\},\;1\leq i
  \leq m$. As a consequence for any $i$ in $[1 .. m]$
  $\lambda \times w_i \in \mathbb{N}$, $\lambda\times c_i \in \mathbb{N}$. Now
  we have to choose the precision which allows us to stop our binary
  search. For this, we take a look at the possible finish times of the workers: all of them are
linear combinations of the different $c_i$ and $w_i$-values. So if we multiply
  all values with $\lambda$ we get integers for all values and the smallest gap
  between two finish times is at least 1. So the precision $p$, i.e., the minimal gap
  between two feasible finish times, is $p = \frac{1}{\lambda}$.\\

\paragraph{Complexity:}
The maximal number of different values $M$ we have to try can be computed as
follows: we examine our algorithm in the interval $[\min(f_i) .. \max(f_i)]$. The possible values have an increment of
$\frac{1}{\lambda}$. So there are $(\max(f_i) - \min(f_i)) \times \lambda$ possible values for $M$.

So the complexity of the binary search is $O(log((\max(f_i)-\min(f_i))\times \lambda))$. Now we
  have to prove that we stay in the order of the size of our problem.
Our platform parameters $c$ and
  $w_i$ are given in the form $w_i = \frac{\alpha_i}{\beta_i}$ and $c =
  \frac{\gamma_i}{\delta}$. So it takes $\log(\alpha_i) + \log(\beta_i)$ to
  store a $w_i$ and $\log(\gamma) + \log(\delta)$ to store a $c$. So our
  entry $E$ has the following size:
$$E = {\displaystyle \sum_{i} \log(\alpha_i) + \sum_i \log(\beta_i) +
  \log(\gamma) +  \log(\delta) + \sum_i \log(L_i)}$$
We can do the following estimation:
$$E {\displaystyle\geq \sum_i \log(\beta_i) +  \log(\delta)} =
\log\left(\prod_{i} \beta_i \times \delta\right) \geq \log(\lambda)$$
So we already know that our complexity is bounded by $O(|E| + \log(\max(f_i)-\min(f_i)))$. We
can simplify this expression: $O(|E| + \log(\max(f_i)-\min(f_i))) \leq O(|E| + \log(\max(f_i)))$.
It remains to upperbound $\log(\max(f_i))$.

Remember $\max(f_i)$ is defined as $\max(f_i) = \max_i(L_i\times w_i) = L_{i_{0}} \times
w_{i_{0}}$. Thus $\log(max(f_i)) = \log (L_{i_0}) + \log(w_{i_0})$. $L_{i_0}$ is a
part of the input and hence its size can be upper-bounded by the size of
the input $E$.
In the same manner we can upperbound $\log(w_{i_0})$ by $\log(w_{i_0}) =
\log(\alpha_{i_0}) + \log(\beta_{i_0}) \leq E$.

Assembling all these upperbounds, we get $O(\log((\max(f_i)-\min(f_i))\times \lambda)) \leq O(3|E|)$
and hence our proposed algorithm needs $O(|E|)$ steps for the binary search. The
total complexity finally is $O(|E| \times \max(nm,\,n^2))$, where $n$ is the
number of scheduled tasks and m the number of workers.

\end{proof}

\subsection{Heuristics for heterogeneous platforms}
\label{sec:heuristic}
As there exists no optimal algorithm to build a schedule in polynomial runtime (unless P
= NP) for heterogeneous
platforms, we propose three heuristics. A comparative study
is done in Section~\ref{sec:tests}.

\begin{itemize}
\item
  The first heuristic consists in the use of the optimal algorithm for
  homogeneous platforms \greedy (see Algorithm~\ref{algo:star_hom}). On heterogeneous platforms, at each step
  \greedy optimizes the local makespan.

\item
  Another heuristic is the utilization of the optimal algorithm for platforms with homogeneous communication
  links \moore (see Algorithm~\ref{algo:star_het}).
  The reason why \moore is not
  optimal on heterogeneous platforms is the following: Moore's algorithm, that
  is used for the scheduling step, cares about the tasks already on the master,
  but it does not assert if the tasks have already arrived. The use of homogeneous
  communication links eliminated this difficulty. We can observe that in the cases where the
  overcharged workers (i.e., the senders) communicate faster than the
  undercharged workers (i.e., the receivers), \moore
  is also optimal. However, the problem with
  this statement is that we do not know a priori which processors will work
  as senders. So in the case of heterogeneous platforms,
  where sending workers have faster communication links than receiving ones, the results will be optimal.

  \item
    We propose a third heuristic: the \backwardalgo (see Algorithm~\ref{algo:backward} for details). This algorithm copies
    the idea of the introduction of deadlines. Contrary to \moore this algorithm
    traverses the deadlines in reversed order, wherefrom the name.
    Starting at a given makespan, \backward schedules all tasks as late
    as possible until no more task can be scheduled.

    \paragraph{\backward can be divided into four phases:}

    \begin{description}
    \item[Phase 1] is the same as in \moore. It decides which of the workers will be senders and which receivers, depending
      of the given makespan (see Figure~\ref{fig:initialDist}).

    \item[Phase 2] fixes how many transfers have to be scheduled from each sender
      such that the senders all finish their remaining tasks in time. This phase
      is also identical to \moore.

    \item[Phase 3] computes for each receiver at which time it can start with the
      computation of the additional tasks, this is in general the given makespan.

    \item[Phase 4] again is the proper scheduling step: Beginning at the makespan
      we fill backward the idle times of the receiving workers. So the master decides which tasks have to
      be scheduled on which receivers and in which order. The master chooses a worker that can start to receive the task as late as possible and
      still finish it in time.

    \end{description}

  \end{itemize}
  \begin{algorithm}[htbp]
    \caption{\backwardalgo}
    \label{algo:backward}
    \begin{algorithmic}[1]

      \STATE {\small /* Phase 1: Initialization */}
      \STATE $T \gets M$; $L_{sched} \gets 0$; $\sigma \gets \emptyset$
      \STATE $\forall{k}\;L_{k}^{(0)} \gets L_{k}$
      \STATE  initialize $end_{i}$ for all workers $i$: $end_{i} = L_{i} \times
      w_{i}$
      \STATE compute $R$ and $S$, order $S$ by non-decreasing values $c_{i}$:
      $c_{s_{1}} \leq c_{s_{2}} \leq \dots$
      \STATE $master\_in \gets c_{s_1}$
      \STATE {\small /* Phase 2: Preparing the senders */}
      \FORALL{$s_{i} \in S$}
      \STATE $l_{s_{i}} \gets \left\lceil \frac{end_{s_{i}} -
          T}{w_{s_{i}}} \right\rceil$
      \IF{$\left\lfloor \frac{T}{c_{s_{i}}}\right\rfloor < l_{s_{i}}$}
    \STATE /* $M$ too small */
    \STATE return ($false, \emptyset$)
    \ENDIF
    \ENDFOR
    \STATE total number of tasks to be sent: $L_{send} \gets \sum_{s_{i}}l_{s_{i}}$

    \STATE {\small /* Phase 3: Determination of the last deadline */}
    \FORALL{$r_i \in R$}
    \IF{$end_{r_i} \leq T$}
    \STATE $begin_{r_i} \gets T$
    \ENDIF
    \ENDFOR

    \STATE {\small /* Phase 4: The scheduling */}
    \WHILE{true}
    \STATE choose $receiver$  such that it is the worker that can start
    receiving it as
    late as possible, i.e., $\max_i\left(\min(begin_i - w_i, T)\right) -
    c_i$ is maximal and that the schedule is feasible: the task must fit in the
    idle gap of the worker: $(begin_{receiver} - w_{receiver} \geq end_{receiver})$ and the task
    has to be arrived at the master: $(begin_{receiver} -
    w_{receiver} - c_{receiver} \geq master\_in)$.
    \IF{no $receiver'$ found}
    \STATE return $((L_{sched}\leq L_{send}),\sigma)$
    \ENDIF
    \STATE $begin_{receiver} \gets begin_{receiver} - w_{receiver}$
    \STATE $T \gets begin_{receiver}-c_{receiver}$
    \STATE $L_{sched} \gets L_{sched} +1$
    \STATE $\sigma \gets \sigma \cup \{receiver\}$

    \STATE $i \gets i+1$
    \ENDWHILE

  \end{algorithmic}
\end{algorithm}

\section{Simulations}
\label{sec:tests}
In this section we present the results of our simulation experiences of the
presented algorithms and heuristics on multiple platforms. We study the
heuristics that we presented in Section~\ref{sec:heuristic}.

\subsection{The simulations}
All simulations were made with \textsc{SimGrid} \cite{legrand_ccgrid03, simgrid}. SimGrid is a
toolkit that provides several functionalities for the simulation of distributed
applications in heterogeneous distributed environments. The toolkit is distributed into
several layers and offers several programming environments, such as XBT, the
core toolbox of SimGrid or
SMPI, a library to run MPI applications on top of a virtual environment. The access to the different components is
ensured via Application Programming Interfaces (API). We use the module MSG to
create our entities.

The simulations were made on automatically created random platforms of four
types: We analyze the behavior on fully homogeneous and fully
heterogeneous platforms and the mixture of both, i.e., platforms with
homogeneous communication links and heterogeneous workers and the converse. For every
platform type 1000 instances were created with the following characteristics:
In absolute random platforms, the random values for $c_i$ and $w_i$ vary between 1 and 100, whereas the
number of tasks is at least 50. In another test series we make some
constraints on the communication and computation powers. In the first one, we
decide the communication power to be inferior to the computation power. In
this case the values for the communication power vary between 20 and 50 and
the computation powers can take values between 50 and 80. In the opposite case, where  communication power is supposed to be superior to the computation
power, these rates are conversed.

\subsection{Trace tests}
To verify the right behavior of the algorithms, we made some trace tests. So
the visualization of the runs on a small test platform are shown in this
section.

We use a small platform with homogeneous communication links, $c = 2$, so the
bandwidth is $0.5$. We use four heterogeneous workers with the following
$w$-values: $P_1$ and $P_2$ compute faster, so we set $w_1 = w_2 = 3$. Worker
$P_3$ and $P_4$ are slower ones with $w_3 = w_4 = 4$. $P_1$ owns 8 tasks at
the beginning, $P_2$ and $P_3$ respectively one task, whereas worker $P_4$ has
no initial work. The optimal makespan is $M=13$, as we computed by permutation
over all possible schedules.

In the following figures, computation are indicated in black. White
rectangles denote internal blockings of SimGrid in the communication
process of a worker. These blockings appear when communication processes
remark that the actual message is not destined for them. Grey
rectangles represent idle time in the computation process. The light grey fields finally show the communication
processes between the processors.

\begin{figure}[htb]
  \centering
  \includegraphics[width=0.9\textwidth]{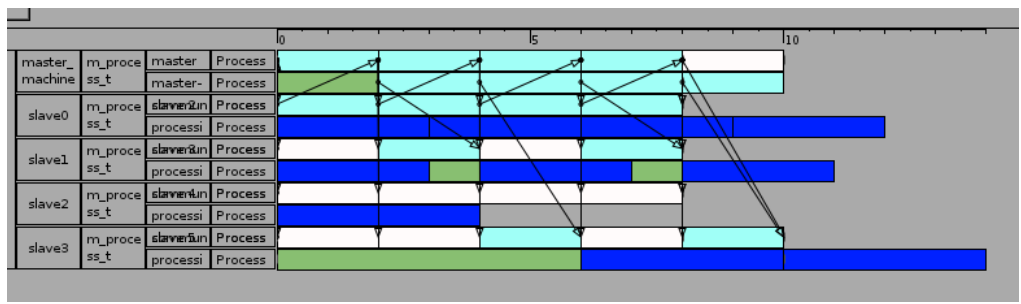}
  \caption{Trace of the simulation of \greedy.}
  \label{fig:greedy}
\end{figure}

The schedule of \greedy can be seen in Figure~\ref{fig:greedy}. Evidently the
worker with the latest finish time is $P_1$, worker $P_2$ can finish the first
sent task earlier than workers $P_3$ and $P_4$, so it is the receiver for the
first task. In this solution, worker $P_1$ sends four tasks, which are
received by $P_2$, $P_4$, $P_2$ and once again $P_4$. The makespan is 14, so
the schedule is not optimal. This does not contradict our theoretical
results, as we proved optimality of \greedy only on homogeneous platforms.

\begin{figure}[htb]
  \centering
  \includegraphics[width=0.9\textwidth]{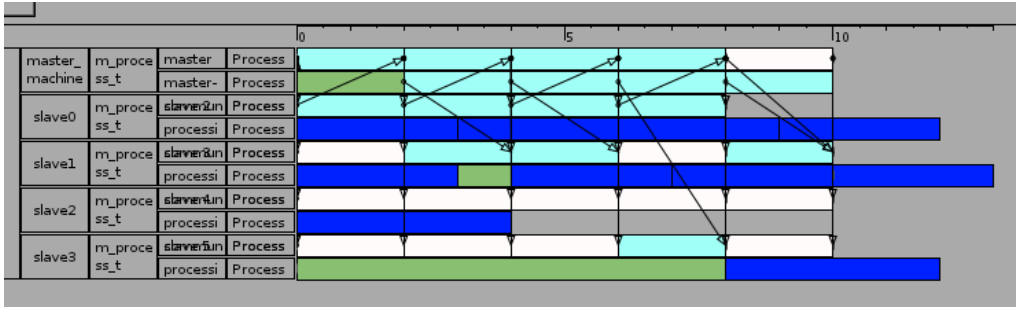}
  \caption{Trace of the simulation of \moore.}
  \label{fig:moore}
\end{figure}

\moore achieves as expected the optimal makespan of 13 (see Figure
\ref{fig:moore}). As you can see by comparing Figures~\ref{fig:greedy} and
\ref{fig:moore}, the second task scheduled by \moore (to worker $P_2$) is
finished processing later than in the schedule of \greedy. So \moore, while
globally optimal, does not minimize the completion time of each task.

\begin{figure}[htb]
  \centering
  \includegraphics[width=0.9\textwidth]{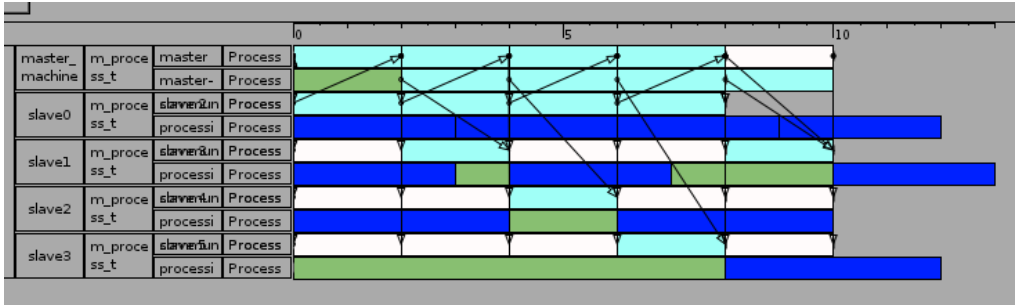}
  \caption{Trace of the simulation of \backward.}
  \label{fig:backward}
\end{figure}

\backward finds also an optimal schedule (cf. Figure~\ref{fig:backward}). Even in this small test the
difference of \backward and \moore is remarkable: \backward tries to schedule
the most tasks as possible by filling idle times starting at the
makespan. \moore contrarily tries to schedule tasks as soon as possible before
their deadlines are expired.

\subsection{Distance from the best}
\label{sec:distance_best}
\begin{figure}[htbp]
    \centering
    \subfigure[Homogeneous platform (general case).]{
      \includegraphics[angle=270,width=0.5\textwidth]{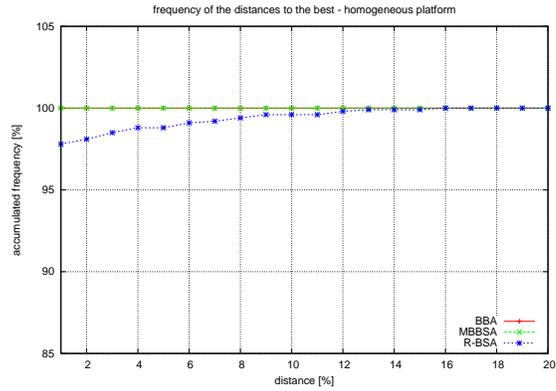}%
      \label{fig:hom}%
    } $\quad$
    \subfigure[Homogeneous platform, faster communicating.]{
      \includegraphics[angle=270,width=0.5\textwidth]{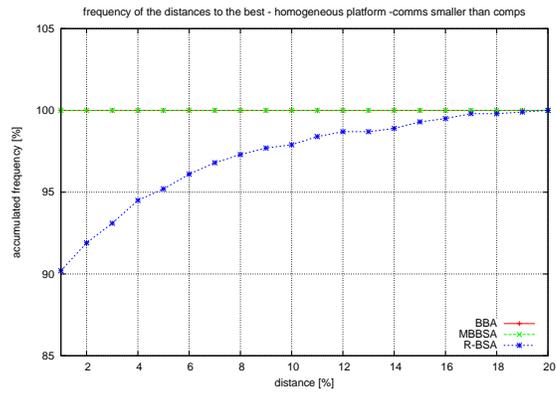}%
      \label{fig:hom_c_kleiner}%
    } $\quad$
    \subfigure[Homogeneous platform, faster computing.]{%
      \includegraphics[angle=270,width=0.5\textwidth]{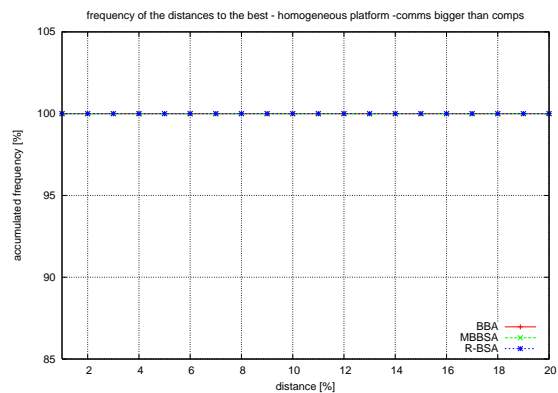}%
      \label{fig:hom_c_groesser}%
    }
    \caption{Frequency of the distance to the best on homogeneous platforms.}
    \label{fig:homogeneous_platform}
  \end{figure}

We made a series of distance tests to get some information of the mean
qualitiy of our algorithms. For this purpose we ran all algorithms on 1000 different random
platforms of the each type, i.e., homogeneous and heterogeneous, as well as
homogeneous communication links with heterogeneous workers and the
converse. We normalized the measured schedule makespans over the best result
for a given instance. In the
following figures we plot the accumulated number of platforms that have a
normalized distance less than the indicated distance. This means, we count on
how many platforms a certain algorithm achieves results that do not differ
more than X\% from the best schedule. For exemple in Figure
\ref{fig:hom_c_kleiner}: The third point of the \backward-line significates that about
93\% of the schedules of \backward differ less than 3\% from the best schedule.

Our results on homogeneous platforms can be seen in Figures~\ref{fig:homogeneous_platform}. As expected from the
theoretical results, \greedy and
\moore achieve the same results and behave equally well on all
platforms. \backward in contrast shows a sensibility on the platform
characteristics. When the communication power is less than the computation
power, i.e. the $c_i$-values are bigger, \backward behaves as good as \moore
and \greedy. But in the case of small $c_i$-values or on homogeneous platforms
without constraints on the power rates, \backward achieves worse results.

\begin{figure}[htbp]
    \centering
    \subfigure[General platform.]{
      \includegraphics[angle=270,width=0.5\textwidth]{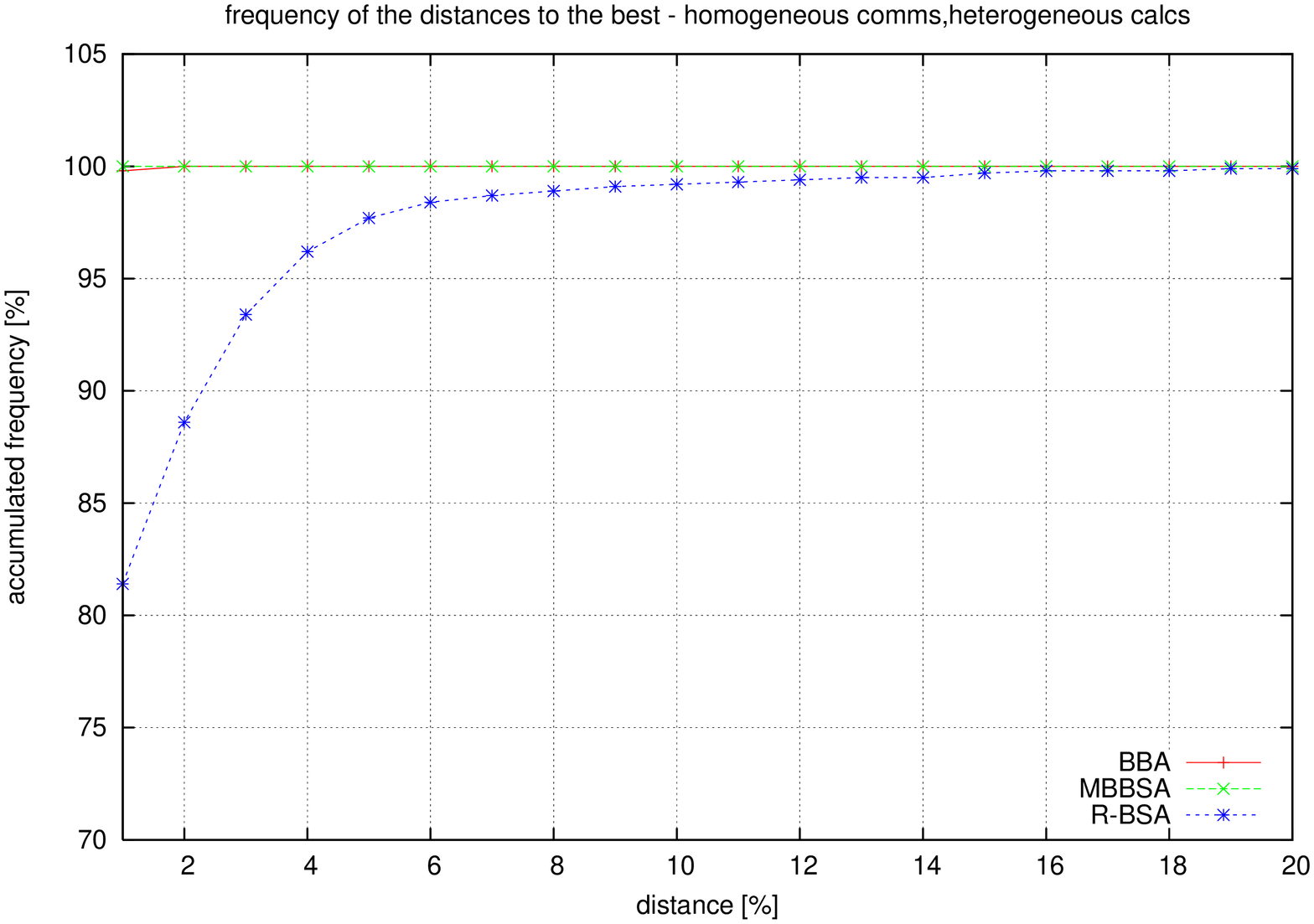}%
      \label{fig:hom_het}%
    } $\quad$
    \subfigure[Faster communicating.]{
      \includegraphics[angle=270,width=0.5\textwidth]{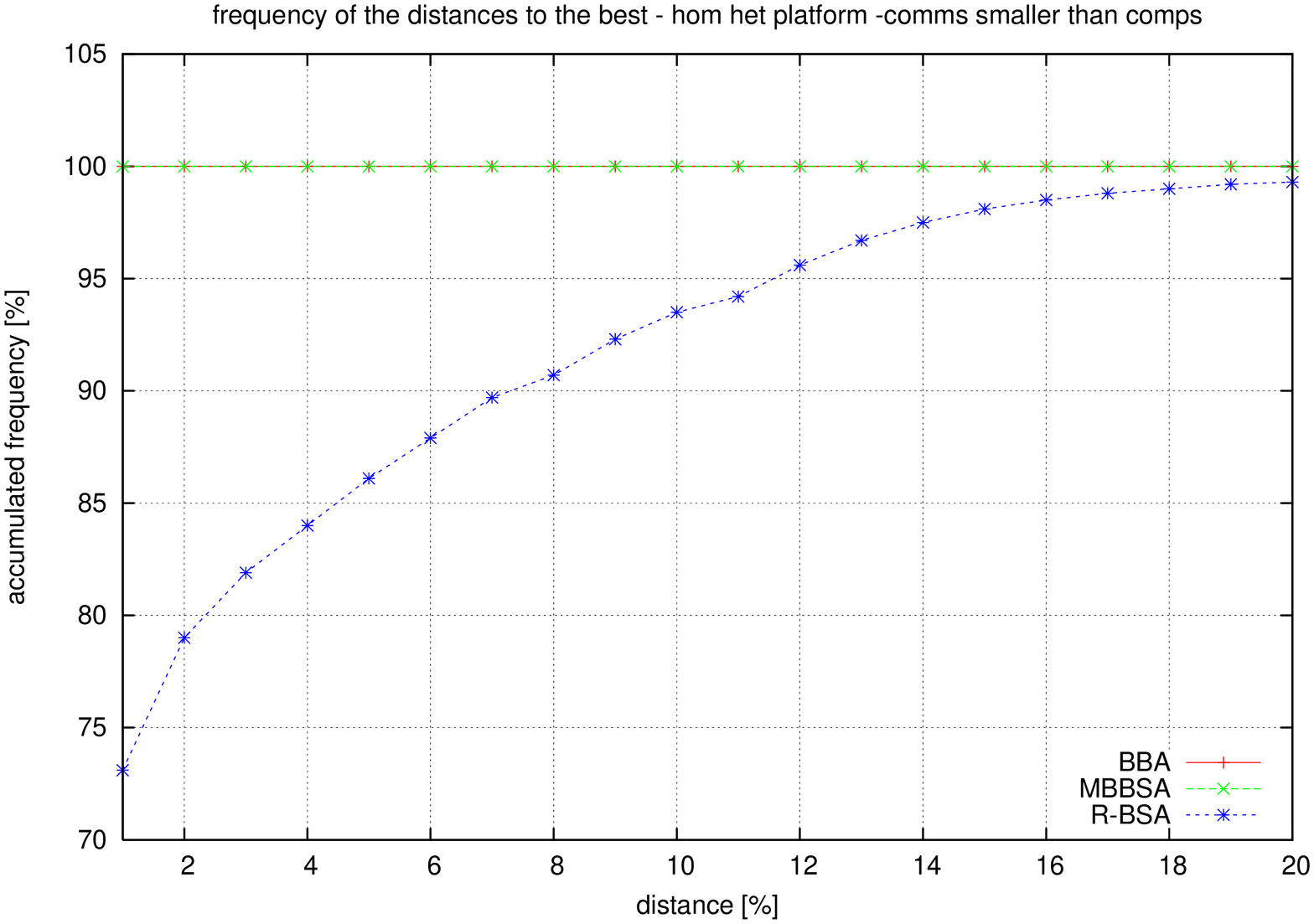}%
      \label{fig:hom_het_c_kleiner}%
    } $\quad$
    \subfigure[Faster computing.]{%
      \includegraphics[angle=270,width=0.5\textwidth]{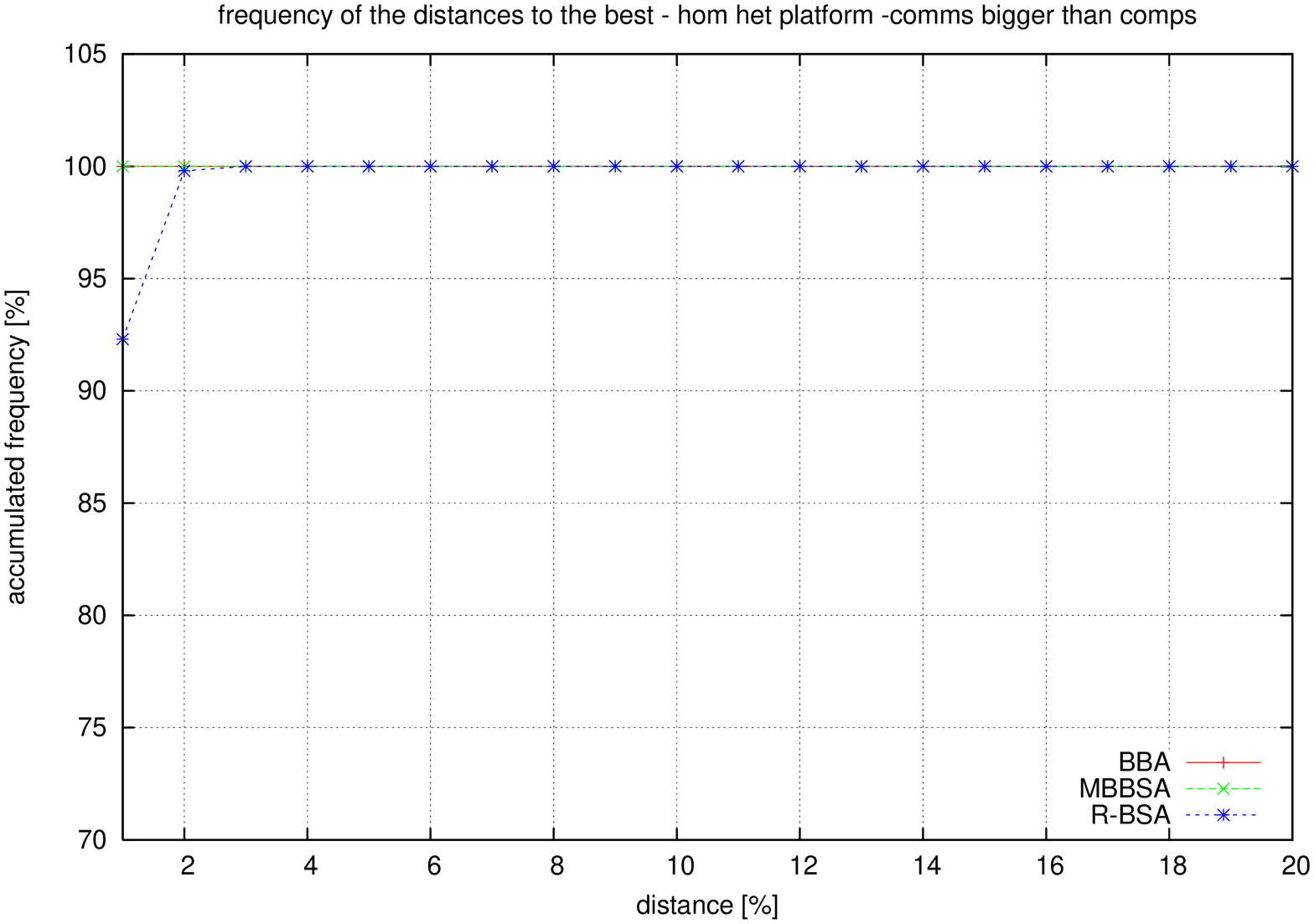}%
      \label{fig:hom_het_c_groesser}%
    }
    \caption{Frequency of the distance to the best on platforms with
      homogeneous communication links and heterogeneous computation power.}
    \label{fig:hom_het_platform}
  \end{figure}

The simulation results on platforms with homogeneous communication links and
heterogeneous computation powers (cf. Figure~\ref{fig:hom_het_platform}) consolidate the theoretical predictions:
Independently of the platform parameters, \moore always obtains optimal
results, \greedy differs slightly when high precision is demanded. The behavior of
\backward strongly depends on the platform parameters: when communications are
slower than computations, it achieves good results.

\begin{figure}[htbp]
    \centering
    \subfigure[General platform.]{
      \includegraphics[angle=270,width=0.5\textwidth]{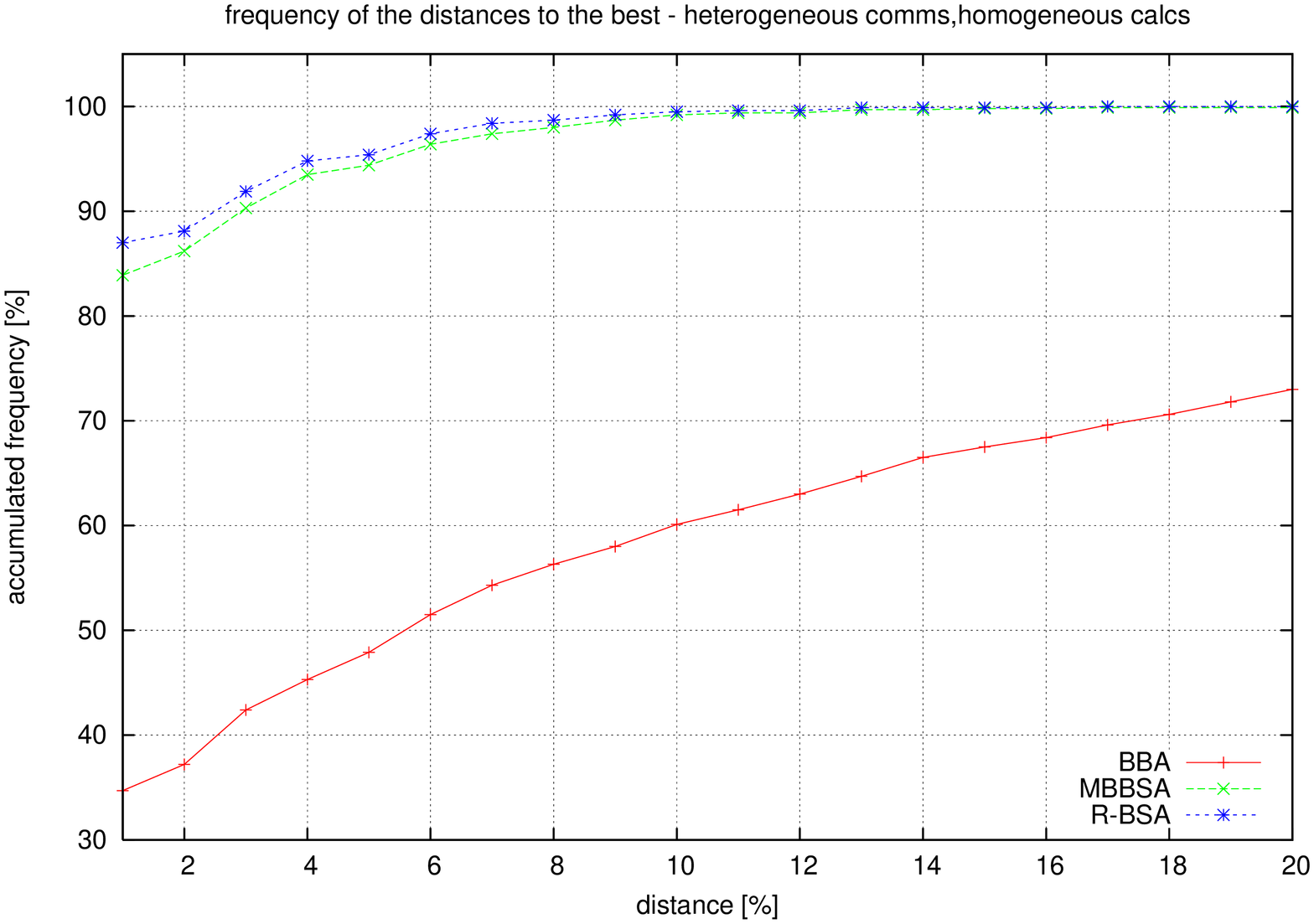}%
      \label{fig:het_hom}%
    } $\quad$
    \subfigure[Faster communicating.]{
      \includegraphics[angle=270,width=0.5\textwidth]{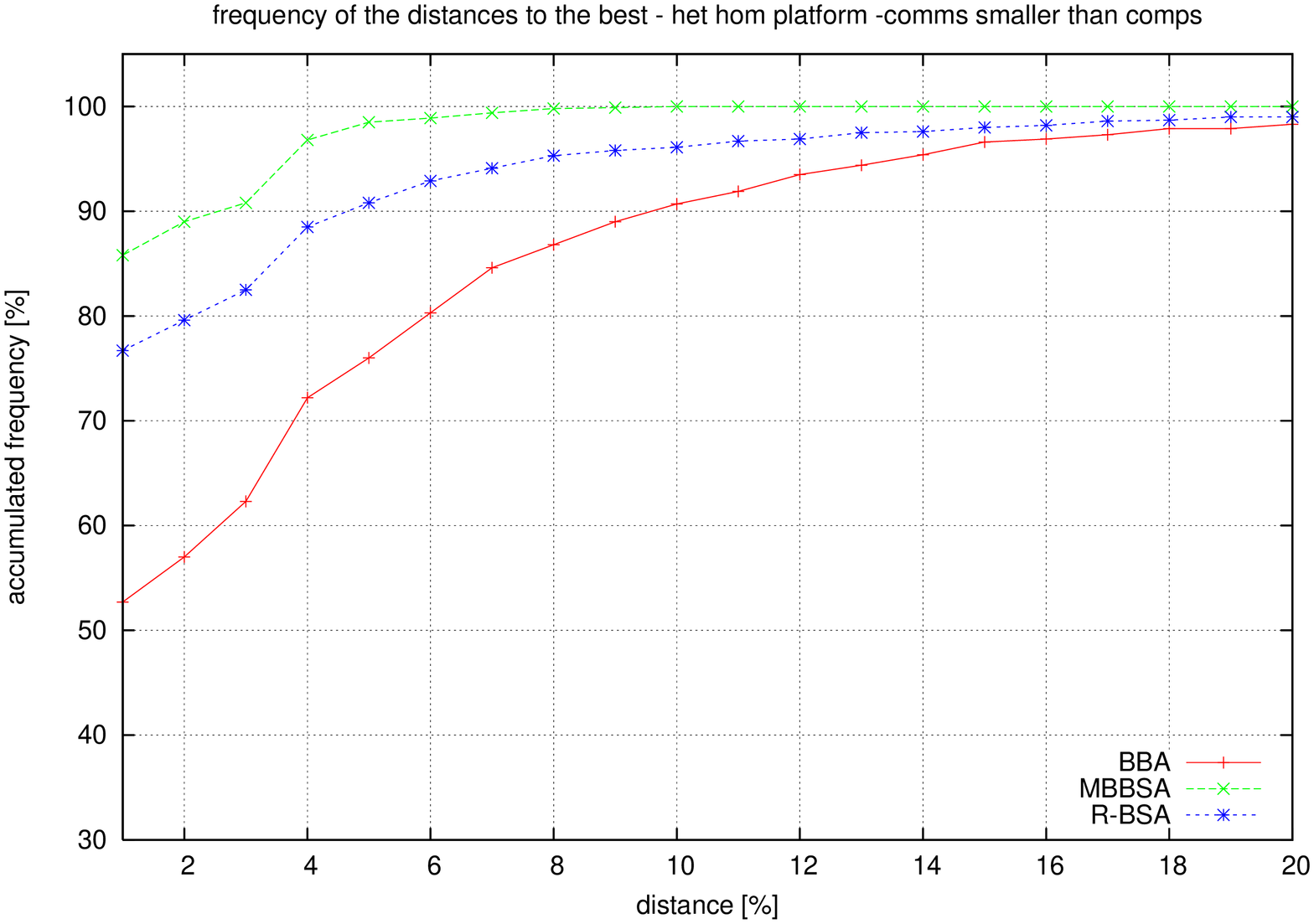}%
      \label{fig:het_hom_c_kleiner}%
    } $\quad$
    \subfigure[Faster computing.]{%
      \includegraphics[angle=270,width=0.5\textwidth]{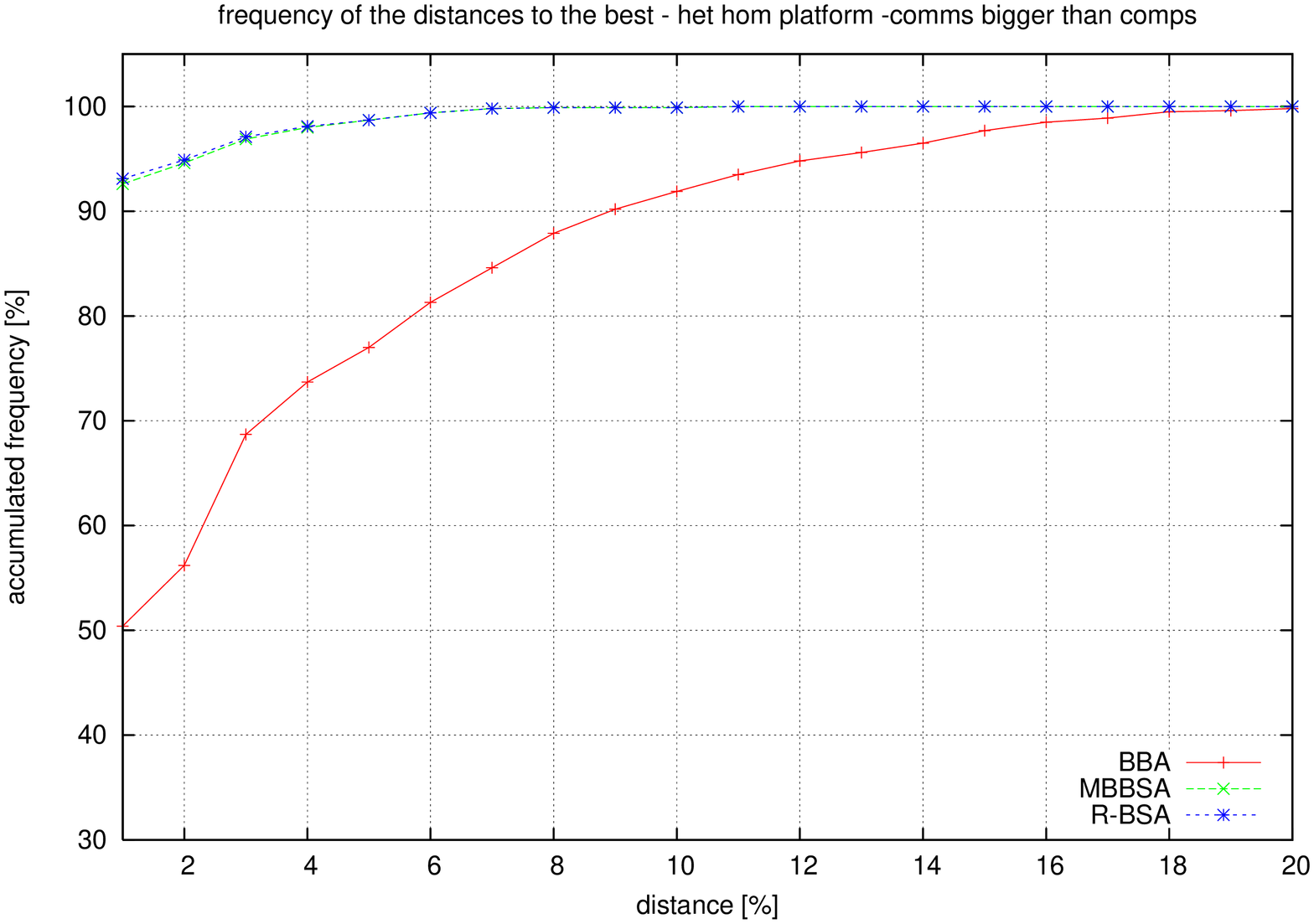}%
      \label{fig:het_hom_c_groesser}%
    }
    \caption{Frequency of the distance to the best on platforms with
      heterogeneous communication links and homogeneous computation power.}
    \label{fig:het_hom_platform}
  \end{figure}

On platforms with heterogeneous communication links and homogeneous
workers, \greedy has by far the poorest results, whereas \backward shows a
good behavior (see Figure~\ref{fig:het_hom_platform}). In general it outperforms \moore, but when the communication
links are fast, \moore is the best.

\begin{figure}[htbp]
    \centering
    \subfigure[Heterogeneous platform (general case).]{
      \includegraphics[angle=270,width=0.5\textwidth]{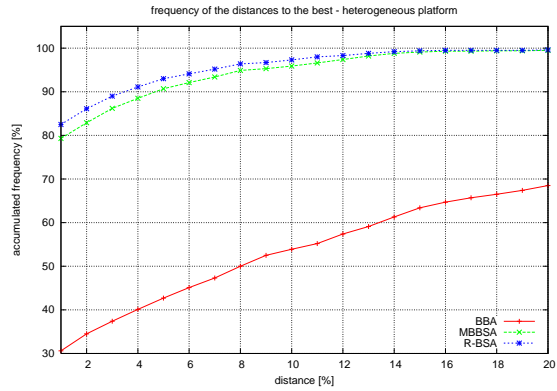}%
      \label{fig:het}%
    } $\quad$
    \subfigure[Heterogeneous platform, faster communicating.]{
      \includegraphics[angle=270,width=0.5\textwidth]{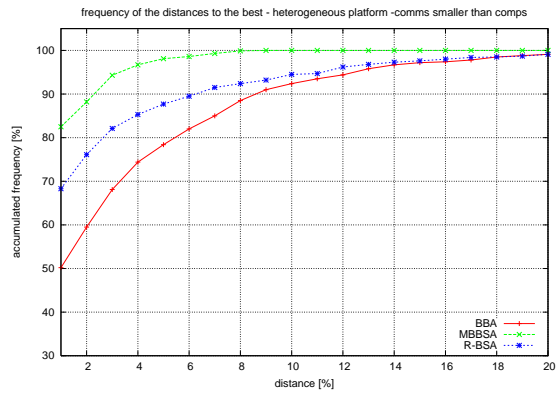}%
      \label{fig:het_c_kleiner}%
    } $\quad$
    \subfigure[Heterogeneous platform, faster computing.]{%
      \includegraphics[angle=270,width=0.5\textwidth]{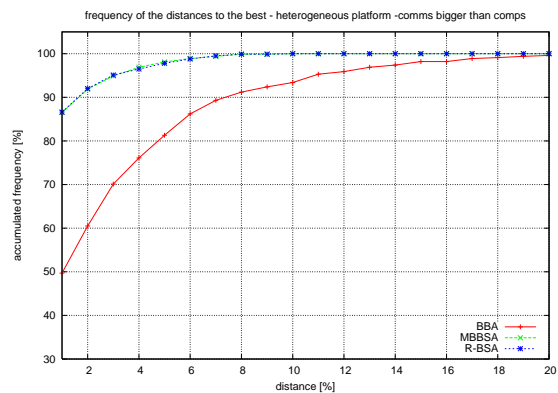}%
      \label{fig:het_c_groesser}%
    }
    \caption{Frequency of the distance to the best on heterogeneous platforms.}
    \label{fig:heterogeneous_platform}
  \end{figure}

The results on heterogeneous platforms are equivalent to these on platforms with heterogeneous communication links and homogeneous
workers, as can be seen in Figure~\ref{fig:heterogeneous_platform}. \backward seems to be a good candidate, whereas \greedy is to avoid as
the gap is up to more than 40\%.

\subsection{Mean distance and standard deviation}
We also computed for every algorithm the mean distance from the best on each
platform type. These calculations are based on the simulation results on the
1000 random platforms of Section~\ref{sec:distance_best}.
As you can see in Table~\ref{tab:mean} in general \moore achieves the best
results. On homogeneous platforms \greedy behaves just as well as \moore and
on platforms with homogeneous communication links it also performs as
well. When communication links are heterogeneous and there is no knowledge
about platform parameters, \backward outperforms the other algorithms and
\greedy is by far the worse choice.
\begin{table}[hbt!]
  \centering
  \begin{tabular}{c|c|c|c c c| c c c}
    \hline
    \multicolumn{3}{c}{Platform type} & \multicolumn{3}{|c|}{Mean distance} &
    \multicolumn{3}{c}{Standard deviation}\\
    Comm. & Comp.& & \greedy & \moore & \backward & \greedy & \moore & \backward \\
    \hline
    Hom & Hom & & 1 & 1 & 1.0014  & 0 & 0 & 0.0107\\
    Hom & Hom & $c\leq w$ & 1 & 1 & 1.0061 & 0 & 0 & 0.0234\\
    Hom & Hom & $c \geq w$&  1 & 1 & 1 & 0 & 0 & 0\\
    \hline
    Hom & Het& &  1.0000 & 1 & 1.0068 &  0.0006 & 0 & 0.0181\\
    Hom & Het& $c\leq w$& 1.0003 & 1 & 1.0186 & 0.0010 & 0 & 0.0395\\
    Hom & Het &$c \geq w$& 1 & 1 & 1.0017 & 0 & 0 & 0.0040\\
    \hline
    Het &Hom &&  1.1894 & 1.0074 & 1.0058 &  0.4007 & 0.0208 & 0.0173\\
    Het &Hom &$c\leq w$& 1.0318 & 1.0049 & 1.0145 & 0.0483 & 0.0131 & 0.0369\\
    Het &Hom &$c \geq w$& 1.0291 & 1.0025 & 1.0024 & 0.0415 & 0.0097 & 0.0095\\
    \hline
    Het &Het& & 1.2100 & 1.0127 & 1.0099 & 0.3516 & 0.0327 & 0.0284\\
    Het &Het & $c\leq w$&1.0296 & 1.0055 & 1.0189 &  0.0450 & 0.0127 & 0.0407\\
    Het & Het & $c \geq w$& 1.0261 & 1.0045 & 1.0046 & 0.0384 & 0.0118 & 0.0121\\
    \hline
  \end{tabular}
  \caption{Mean distance from the best and standard deviation of the different algorithms on each platform type.}
  \label{tab:mean}
\end{table}

The standard deviations of all algorithms over the 1000 platforms are shown in
the right part of
Table~\ref{tab:mean}. These values mirror exactly the same conclusions as
the listing of the mean distances in the left part, so we do not
comment on them particularly. We only want to point out that the standard
deviation of \moore always keeps small values, whereas in case of
heterogeneous communication links \greedy-heuristic is not recommendable.

\section{Load balancing of divisible loads using the multiport switch-model}
\label{sec:divisible}

\subsection{Framework}
\label{sec:framework_divisible}

In this section we work with a heterogeneous star network. But in difference to
Section~\ref{sec:tasks} we replace the master by a switch. So we have $m$
workers which are interconnected by a switch and $m$ heterogenous links.
Link $i$ is the link that connects worker $P_i$ to the switch. Its
bandwidth is denoted by $b_i$. In the same way $s_i$ denotes the computation
speed of worker $P_i$. Every worker $P_i$ possesses an amount of initial load
$\alpha_i$. Contrarily to the previous section, this load is not considered to
consist of identical and independent tasks but of divisible loads. This means that
an amount of load $X$ can be divided into an arbitrary number of tasks of
arbitrary size. As already mentioned, this approach is called \emph{Divisible Load Theory - DLT}  \cite{Bharadwaj-Cluster2003}.
The communication model used in this case is an overlapped unbounded switched-multiport
model. This means all communications pass by a centralized switch that has no
throughput limitations. So all workers can communicate at the same time and a
given worker can start executing as soon as it receives the first bit of data. As we use a model with overlap, communication and computation can take place
at the same time.

As
in the previous section our objective is to balance the load over the participating
workers to minimize the global
makespan $M$.

\subsection{Redistribution strategy}
Let $\sigma$ be a solution of our problem that takes a time $T$. In this
solution, there is a set of sending workers $S$ and a set of receiving workers
$R$. Let $send_i$ denote the amount of load sent by sender $P_i$ and
$recv_j$ be the amount of load received by receiver $P_j$, with $send_i \geq 0$,
$recv_j \geq 0$. As all load that is sent has to be received by another worker, we
have the following equation:
\begin{equation}\label{eq:balance}
\sum_{i \in S}send_i = \sum_{j \in R}recv_j = L.
\end{equation}
In the following we describe the properties of the senders: As the solution
$\sigma$ takes a time $T$, the amount of load a sender can send depends on its bandwidth:
So it is bounded by the time-slot of
\begin{equation}\label{eq:sender}
\forall\; \textnormal{sender}_i\in S,\hspace{1cm} \frac{send_i}{b_i}\leq T.
\end{equation}
Besides, it has to send at least the amount of load that it can not finish
processing in time $T$. This lowerbound can be expressed by
\begin{equation}\label{eq:to_send}
\forall\; \textnormal{sender}_i\in S,\hspace{1cm} send_i \geq \alpha_i -
T\times s_i.
\end{equation}

The properties for receiving workers are similar. The amount of load a worker
can receive is dependent of its bandwidth. So we have:
\begin{equation}\label{eq:receiver}
\forall\; \textnormal{receiver}_j\in R,\hspace{1cm}\frac{recv_j}{b_j}\leq T.
\end{equation}
Additionally it is dependent of the amount of load it already possesses and of
its
computation speed. It must have the time to process all its load, the initial
one plus the received one. That is why we have a second upperbound:
\begin{equation}\label{eq:to_recv}
\forall\; \textnormal{receiver}_j\in S,\hspace{1cm} \frac{\alpha_j + recv_j}{s_j} \leq T.
\end{equation}

For the rest of our paper we introduce a new notation: Let $\delta_i$
denote the imbalance of a worker. We will define it as follows:
$$\delta_i = \begin{cases}
  send_i &   \text{if $i \in S$}\\
  - recv_i&      \text{if $i \in R$}
\end{cases}
$$.

 With the help of this new notation we can re-characterize the imbalance of all workers:

\begin{itemize}

\item
  This imbalance is bounded by $$|\delta_i| \leq b_i \times T.$$
  \begin{itemize}
  \item
    If $i \in S$, worker $P_i$ is a sender, and this statement is true
    because of inequality~\ref{eq:sender}.
  \item
    If $i \in R$, worker $P_i$ is a receiver
    and the statement is true as well, because of inequality~\ref{eq:receiver}.
  \end{itemize}

\item
  Furthermore, we lower-bound the imbalance of a worker by
  \begin{equation}\label{eq:lowerbound}
    \delta_i \geq \alpha_i - T\times s_i.
  \end{equation}
  \begin{itemize}
  \item
    If $i \in S$, we are in the case where $\delta_i = send_i$ and hence this
    it true because of equation~\ref{eq:to_send}.
  \item
    If $i \in R$, we have $\delta_i = -recv_i \leq 0$. Hence we get that (\ref{eq:lowerbound})
    is equal to $-recv_i \geq \alpha_i - T\times s_i$
    which in turn is equivalent to (\ref{eq:to_recv}).
  \end{itemize}
\item
  Finally we know as well that $\sum_{i}\delta_i = 0$ because of equation~\ref{eq:balance}.
\end{itemize}

If we combine all these constraints we get the following linear program (LP),
with the addition of our objective to minimize the makespan $T$.
This combination of all properties into a LP is possible because we can use the same
constraints for senders and receivers. As you may have noticed, a worker will
have the functionality of a sender if its imbalance $\delta_i$ is positive,
receivers being characterized by negative $\delta_i$-values.

\begin{LinearProgram}{T}
  \label{min_makespan}\ensuremath
  \EquationsNumbered{\mark
    |\delta_i| \leq T\times b_i\n%
    \delta_i \geq \alpha_i - T\times s_i\n%
    \sum_i \delta_i =  0
  }
\end{LinearProgram}
All the constraints of the LP are satisfied for the $(\delta_i,T)$-values of
any schedule solution of the initial problem.
We call $T_0$ the solution of the LP for a given problem. As the LP minimizes
the time $T$, we have $T_0 \leq T$ for all valid schedule and hence we have
found a lower-bound for the optimal
makespan.

Now we prove that we can find a feasible schedule with makespan $T_0$. We
start from an optimal solution of the LP, i.e., $T_0$ and the
$\delta_i$-values computed by some LP solvers, such as Maple or MuPAD.
With the help of these found values we are able to describe the schedule:
\begin{enumerate}
\item Every sender $i$ sends a fraction of load to each receiver
  $j$.
We decide that each sender sends to each receiver a fraction of the senders
  load proportional to what we denote by
\begin{equation}\label{eq:fij}
    f_{i,j} = \delta_i\times \frac{\delta_j}{\sum_{k\in R}
      \delta_k} =\delta_i\times \frac{\delta_j}{-L}
  \end{equation}
  the fraction of load that a sender $P_i$ sends to a
  receiver $P_j$. In other words we have
  $f_{i,j} = \delta_i\times \frac{-recv_i}{\sum_{k\in R}(-recv_k)}$.

\item
  During the whole schedule we use constant communication rates, i.e., worker
  $j$ will receive its fraction of load $f_{i,j}$ from sender $i$ with a fixed
  receiving rate, which is denoted by $\lambda_{i,j}$:
  \begin{equation}\label{eq:lambda}
    \lambda_{i,j} = \frac{f_{i,j}}{T_0}.
  \end{equation}

\item
  A schedule starts at time $t=0$ and ends at time $t= T_0$.

\end{enumerate}

  We have to
  verify that each sender can send its amount of load in time $T_0$ and that
  the receivers can receive it as well and compute it afterwards.

  Let us take a look at a sender $P_i$:
  the total amount it will send is $\sum_{j\in R}f_{i,j} = \sum_{j \in R}
  \frac{\delta_i\times \delta_j}{\sum_{k\in R}\delta_k} = \delta_i = send_i$
  and as we started by a solution of our LP, $\delta_i$ respects equations~\ref{min_makespan}a and~\ref{min_makespan}b, thus $send_i$ respects the constraints~\ref{eq:sender} and~\ref{eq:to_send} as well, i.e., $send_i \leq T \times
  b_i$ and $send_i \geq \alpha_i - T \times s_i$.

\begin{figure}[htbp]
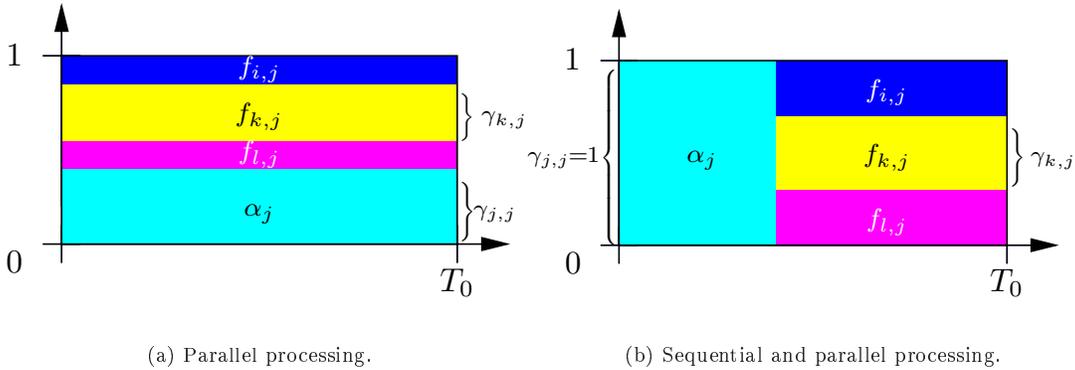

    \centering
    \subfigure[Parallel processing.]{
      \includegraphics[width=0.45\textwidth]{divisibleParallel.fig}%
      \label{fig:computing_parallel}%
    } $\quad$
    \subfigure[Sequential and parallel processing.]{%
      \includegraphics[width=0.45\textwidth]{divisibleSequPara.fig}%
      \label{fig:computing_sequentiel}%
    }
    \caption{Different schedules to process the received load.}
    \label{fig:computing}
  \end{figure}

  Now we
  consider a receiver $P_j$: the total amount it will receive is $\sum_{i\in
    S}f_{i,j} = \sum_{i \in S}\frac{\delta_i\times \delta_j}{\sum_{k\in
      R}\delta_k} = -\delta_j = recv_j$. Worker $P_i$ can receive the whole amount
  of $recv_i$ load in time $T_0$ as it starts the reception at time $t=0$ and
  $recv_i$ respects constraints~\ref{min_makespan}a and~\ref{min_makespan}b, who in turn respect
  the initial constraints~\ref{eq:receiver} and~\ref{eq:to_recv}, i.e.,
  $recv_i \leq T\times b_i$ and $recv_i \leq T\times s_i -\alpha_i$. Now we examine if worker $P_i$ can
  finish computing all its work in time. As we use the divisible load model,
  worker $P_i$ can start computing its additional amount of load as soon as it
  has received its first bit and provided the computing rate is inferior to
  the receiving rate.
  Figure~\ref{fig:computing} illustrates the
  computing process of a receiver. There are two possible schedules: the
  worker can allocate a certain percentage of its computing power for each
  stream of loads and process them in parallel.  This is shown in Figure~\ref{fig:computing_parallel}. Processor $P_i$ starts immediately processing
  all incoming load. For doing so, every stream is allocated a
  certain computing rate $\gamma_{i,j}$, where $i$ is the sending worker and $j$
  the receiver. We have to verify that  the computing rate is inferior or
  equal to the receiving rate. 

  The initial load $\alpha_j$ of receiver $P_j$ owns at minimum a computing rate
  such that it finishes right in time $T_0$: $\gamma_{j,j} = \frac{\alpha_j}{T_0}$.
  The computing rate $\gamma_{i,j}$, for all pairs $(i,j)$, $i \in S$, $j \in
  R$, has to verify the following constraints:
  \begin{itemize}
  \item
    The sum of all computing rates does not exceed the computing power
    $s_j$ of the worker $P_j$:
    \begin{equation}\label{eq:one}
      \left(\sum_{i\in S}\gamma_{i,j}\right) + \frac{\alpha_j}{T_0} \leq s_j,
    \end{equation}
  \item
    The computing rate for the amount of load $f_{i,j}$ has to be sufficiently big
    to finish in time $T_0$:
    \begin{equation}\label{eq:two}
      \gamma_{i,j} \geq \frac{f_{i,j}}{T_0},
    \end{equation}
  \item
    The computing rate has to be inferior or equal to the receiving rate of
    the amount $f_{i,j}$:
    \begin{equation}\label{eq:three}
      \gamma_{i,j} \leq \lambda_{i,j},
    \end{equation}
  \end{itemize}

Now we prove that $\gamma_{i,j} = \frac{f_{i,j}}{T_0}$  is a valid solution
that respects constraints (\ref{eq:one}), (\ref{eq:two}), and (\ref{eq:three}):
\begin{description}
\item[Equation (\ref{eq:one})]
  We have $\left(\sum_{i\in S}\gamma_{i,j}\right) + \frac{\alpha_j}{T_0} =
  \left(\sum_{i\in S} \frac{f_{i,j}}{T_0}\right) + \frac{\alpha_j}{T_0}
  =\left( \frac{-\delta_{j}}{T_0}\right) + \frac{\alpha_j}{T_0} =
  \frac{\alpha_j - \delta_j}{T_0}$. Transforming Equation
  (\ref{min_makespan}b) in $\alpha_j - \delta_j \leq T_0 \times s_j$ and using
  this upperbound we get  $\frac{\alpha_j - \delta_j}{T_0} \leq \frac{T_0
  \times s_j}{T_0} = s_j$. Hence this constraint holds true.
\item[Equation (\ref{eq:two})]
  By definition of $\gamma_{i,j}$ this holds true.
\item[Equation (\ref{eq:three})]
  By the definitions of $\gamma_{i,j}$ and $\lambda_{i,j}$ this holds true.
\end{description}

  In the other possible schedule, all incoming load streams are
  processed in parallel after having processed the initial amount of load as shown in Figure~\ref{fig:computing_sequentiel}.  In
  fact, this modeling is equivalent to the precedent one, because we use the
  DLT paradigm. We used
  this model in equations~\ref{eq:to_send} and~\ref{eq:to_recv}.

The following theorem summarizes our cognitions:

\begin{theorem}
The combination of the linear program~\ref{min_makespan} with equations
\ref{eq:fij} and~\ref{eq:lambda} returns an optimal solution for makespan
minimization of a load balancing problem on a heterogeneous star platform using the switch model and initial loads on the workers.
\end{theorem}

\section{Conclusion}
In this report we were interested in the problem of scheduling and
redistributing data on master-slave platforms. We considered two types of
data models.

Supposing independent and identical tasks, we were able to
prove the NP completeness in the strong sense for the general case of completely heterogeneous platforms. Therefore we restricted this case to
the presentation of three heuristics. We have also proved that our problem is
polynomial when computations are negligible.
Treating some special topologies, we were able to
present optimal algorithms for totally homogeneous star-networks and for
platforms with homogeneous communication links and heterogeneous workers. Both
algorithms required a rather complicated proof.

The simulative experiments consolidate our theoretical results of
optimality. On homogeneous platforms, \greedy is to privilege over \moore, as the
complexity is remarkably lower. The tests on heterogeneous platforms show
that \greedy performs rather poorly in comparison to \moore and
\backward. \moore in general achieves the best results, it might be
outperformed by \backward when platform parameters have a certain
constellation, i.e., when workers compute faster than they are communicating.

Dealing with divisible loads as data model, we were able to solve the fully
heterogeneous problem. We presented the combination of a linear program with
simple computation formulas to compute the imbalance in a first step and the
corresponding schedule in a second step.

A natural extension of this work would be the following: for the model with independent tasks,
it would be nice to derive approximation algorithms, i.e., heuristics whose worst-case
is guaranteed within a certain factor to the optimal, for the fully heterogeneous case.
However, it is often the case in scheduling problems for heterogeneous platforms that
approximation ratios contain the quotient of the largest platform parameter
by the smallest one, thereby leading to very pessimistic results in practical situations.

More generally, much work remains to be done along the same lines of load-balancing and
redistributing while computation goes on. We can envision dynamic master-slave platforms
whose characteristics vary over time, or even where new resources are enrolled temporarily in the execution.
We can also deal with more complex interconnection networks, allowing slaves to circumvent the
master and exchange data directly.

\bibliographystyle{abbrv}
\bibliography{biblio}

\end{document}